\RequirePackage{fix-cm} 
\documentclass[a4paper, twoside, reqno, dvips, 12pt]{amsart}
\usepackage{fixltx2e}   

\usepackage{etex}

\usepackage[latin1]{inputenc}
\usepackage[T1]{fontenc}

\usepackage{esint}
\usepackage{dsfont}
\usepackage{xspace}
\usepackage{amsgen}
\usepackage{amsthm}
\usepackage{amssymb}
\usepackage{amsmath}
\usepackage{wasysym}
\usepackage{upgreek}
\usepackage{amsfonts}
\usepackage{stmaryrd}
\usepackage{scalerel}
\usepackage{mathtools}
\usepackage{stackengine}

\usepackage{relsize}
\usepackage{textcomp}
\usepackage[mathcal, mathscr]{euscript}

\usepackage{mathrsfs}
\DeclareMathAlphabet{\mathscrbf}{OMS}{mdugm}{b}{n}

\usepackage{a4wide}

\usepackage[dvipsnames, table]{xcolor}
\definecolor{bckg}{RGB}{20.8, 20.8, 20.8}
\definecolor{oneblue}{rgb}{0.0, 0.0, 0.85}
\definecolor{Lightblue}{RGB}{214, 214, 214}
\definecolor{bluepigment}{rgb}{0.2, 0.2, 0.6}
\definecolor{charcoal}{rgb}{0.21, 0.27, 0.31}
\definecolor{denimblue}{rgb}{0.08, 0.38, 0.74}
\definecolor{Lightgray}{rgb}{0.89, 0.89, 0.89}
\definecolor{darkgrey}{rgb}{0.273, 0.281, 0.30}
\definecolor{darkelectricblue}{rgb}{0.33, 0.41, 0.47}

\usepackage{multirow}

\usepackage[sort&compress, comma, square, numbers]{natbib}

\usepackage{psfrag}
\usepackage{graphicx}
\usepackage{adjustbox}
\usepackage[tight]{subfigure}
\usepackage{morefloats}
\usepackage{indentfirst}

\usepackage[usenames, dvipsnames, pdf]{pstricks}
\usepackage{epsfig}
\usepackage{pst-grad} 
\usepackage{pst-plot} 

\usepackage{rotating}
\usepackage{pdflscape}

\usepackage{acronym}
\usepackage{microtype}
\usepackage[labelsep=period,%
            labelfont={bf,sf,color=bluepigment},%
            justification=raggedright]{caption}

\usepackage[perpage, symbol]{footmisc}

\usepackage[colorlinks,
           urlcolor=oneblue,
           linkcolor=denimblue,
           citecolor=NavyBlue,
           bookmarksopen=false,
           pdfpagemode=UseNone,
           pagebackref]{hyperref}

\usepackage[explicit]{titlesec}

\titleformat{\section}[block]
  {\color{NavyBlue}\Large\sffamily\bfseries}
  {}
  {0.0em}
  {\colorbox{bckg!5}{\strut\parbox{\dimexpr\linewidth-2\fboxsep\relax}{\thesection. #1}}}
  [\vspace*{0.33em}]

\titleformat{name=\section,numberless}[block]
  {\color{NavyBlue}\Large\sffamily\bfseries}
  {}
  {0.0em}
  {\colorbox{bckg!5}{\strut\parbox{\dimexpr\linewidth-2\fboxsep\relax}{#1}}}
  [\vspace*{0.33em}]

\titleformat{\subsection}
  {\color{NavyBlue}\large\sffamily\bfseries}
  {}
  {0.0em}
  {\colorbox{bckg!5}{\parbox{\dimexpr\linewidth-2\fboxsep\relax}{\thesubsection. #1}}}
  [\vspace*{0.33em}]

\titleformat{name=\subsection,numberless}
  {\color{NavyBlue}\Large\sffamily\bfseries}
  {}
  {0em}
  {\colorbox{bckg!5}{\parbox{\dimexpr\linewidth-2\fboxsep\relax}{#1}}}
  [\vspace*{0.33em}]

\titleformat{\subsubsection}
  {\color{bluepigment}\sffamily\normalsize\bfseries}
  {\thesubsubsection}
  {0.5em}
  {#1}
  [\vspace*{0.33em}]

\titleformat{\paragraph}[runin]
  {\color{bluepigment}\sffamily\small\bfseries}
  {}
  {0em}
  {#1}

\titlespacing{\section}{0.0em}{1.5em plus 2pt minus 2pt}%
{1.0em plus 2pt minus 2pt}[0em]
\titlespacing{\subsection}{0.5em}{1.5em plus 2pt minus 2pt}%
{1.0em}[0em]
\titlespacing{\subsubsection}{0.5em}{1.5em plus 2pt minus 2pt}%
{1.0em plus 2pt minus 2pt}[0em]

\usepackage{titletoc}

\contentsmargin{0.5em}
\newlength{\tocsep} 
\titlecontents{section}[\tocsep]
  {\addvspace{10pt}\bfseries\sffamily}
  {\contentslabel[\thecontentslabel]{\tocsep}}
  {}
  {\ \titlerule*[0.75pc]{.}\ \thecontentspage}
  []
\titlecontents{subsection}[\tocsep]
  {\addvspace{8pt}\sffamily}
  {\contentslabel[\thecontentslabel]{\tocsep}}
  {}
  {\ \titlerule*[0.5pc]{.}\ \thecontentspage}
  []
\titlecontents*{subsubsection}[\tocsep]
  {\addvspace{2pt}\footnotesize\sffamily}
  {}
  {}
  {\ \titlerule*[0.35pc]{.}\ \thecontentspage}
  [\\*]

\makeatletter
\def\@setauthors{%
  \begingroup
  \def\thanks{\protect\thanks@warning}%
  \trivlist
  \centering\footnotesize \@topsep30\p@\relax
  \advance\@topsep by -\baselineskip
  \item\relax
  \author@andify\authors
  \def\\{\protect\linebreak}%
  \textsc{\normalsize\textcolor{darkelectricblue}{\authors}}%
  \ifx\@empty\contribs
  \else
    ,\penalty-3 \space \@setcontribs
    \@closetoccontribs
  \fi
  \endtrivlist
  \endgroup
}
\def\@settitle{\begin{center}%
  \baselineskip14\p@\relax
    \bfseries
    \textsc{\Large\textcolor{charcoal}{\@title}}
  \end{center}%
}
\makeatother

\usepackage{enumitem}
\setlist[description]{%
  topsep=30pt,               
  itemsep=5pt,               
  font={\bfseries\sffamily\color{NavyBlue}}, 
}

\usepackage{fancyhdr}
\usepackage{lastpage}

\newcommand*\Title{\textcolor{bluepigment}{Peregrine's system revisited}}
\newcommand*\Authors{\textcolor{bluepigment}{A.~Dur\'an, D.~Dutykh \& D.~Mitsotakis}}
\newcommand*{\plogo}{\textcolor{gray}{{\texttt{arXiv.org} / \textsc{hal}}}} 

\numberwithin{equation}{section}

\newtheorem{remark}{Remark}

\newcommand{\up}[1]{$^{\mathrm{\small\textsf{#1}}}$} 

\newcommand{\A}{\mathds{A}}
\newcommand{\K}{\mathds{K}}
\newcommand{\R}{\mathds{R}}
\newcommand{\Z}{\mathds{Z}}
\newcommand{\vm}{\bar{\/v}}
\newcommand{\Hm}{\bar{\/H}}
\newcommand{\Qm}{\bar{\/Q}}
\newcommand{\C}{\mathcal{C}}
\newcommand{\F}{\mathcal{F}}
\newcommand{\N}{\mathcal{N}}
\newcommand{\X}{\mathcal{X}}
\newcommand{\ud}{\mathrm{d}}
\newcommand{\ue}{\mathrm{e}}
\newcommand{\St}{\mathrm{S}}
\newcommand{\Sp}{\mathbb{S}}
\newcommand{\D}{\mathds{D}}
\newcommand{\E}{\mathds{E}}
\newcommand{\I}{\mathds{I}}
\renewcommand{\L}{\mathbb{L}}
\renewcommand{\O}{\mathcal{O}}

\renewcommand{\geq}{\geqslant}
\newcommand{\eps}{\varepsilon}
\newcommand{\const}{\mathrm{const}}

\newcommand{\cf}{\emph{cf.}\xspace}
\newcommand{\ie}{\emph{i.e.}\xspace}
\newcommand{\eg}{\emph{e.g.}\xspace}

\newcommand{\etal}{\emph{et al.}\xspace}

\renewcommand{\sim}{\thicksim}

\newcommand{\sech}{\mathrm{sech}}

\newcommand{\diag}{\mathop{\mathrm{diag}}}
\newcommand{\sign}{\mathop{\mathrm{sign}}}
\newcommand{\norm}[1]{\lVert\, #1\, \rVert}

\newcommand{\minmod}{\mathop{\mathrm{minmod}}}
\newcommand{\abs}[1]{\left\lvert\, #1\, \right\rvert}
\newcommand{\pd}[2]{\frac{\partial\/ #1}{\partial\/ #2}}
\newcommand{\od}[2]{\frac{\mathrm{d}\/ #1}{\mathrm{d}\/ #2}}

\newcommand{\eqdef}{\ \mathop{\stackrel{\;\mathrm{def}}{:=}\;}\ }

\newcommand{\half}{{\textstyle{1\over2}}}
\newcommand{\frth}{{\textstyle{3\over4}}}
\newcommand{\third}{{\textstyle{1\over3}}}

\begin{document}

\title[\Title]{Peregrine's system revisited}

\author[A.~Dur\'an]{Angel Dur\'an}
\address{\textbf{A.~Dur\'an:} Departamento de Matem\'atica Aplicada, E.T.S.I. Telecomunicaci\'on, Campus Miguel Delibes, Universidad de Valladolid, Paseo de Belen 15, 47011 Valladolid, Spain}
\email{angel@mac.uva.es}
\urladdr{https://www.researchgate.net/profile/Angel\_Duran3/}

\author[D.~Dutykh]{Denys Dutykh$^*$}
\address{\textbf{D.~Dutykh:} Univ. Grenoble Alpes, Univ. Savoie Mont Blanc, CNRS, LAMA, 73000 Chamb\'ery, France and LAMA, UMR 5127 CNRS, Universit\'e Savoie Mont Blanc, Campus Scientifique, F-73376 Le Bourget-du-Lac Cedex, France}
\email{Denys.Dutykh@univ-smb.fr}
\urladdr{http://www.denys-dutykh.com/}
\thanks{$^*$ Corresponding author}

\author[D.~Mitsotakis]{Dimitrios Mitsotakis}
\address{\textbf{D.~Mitsotakis:} Victoria University of Wellington, School of Mathematics, Statistics and Operations Research, PO Box 600, Wellington 6140, New Zealand}
\email{dmitsot@gmail.com}
\urladdr{http://dmitsot.googlepages.com/}



\begin{titlepage}
\thispagestyle{empty} 
\noindent
{\Large Angel \textsc{Dur\'an}}\\
{\it\textcolor{gray}{Universidad de Valladolid, Spain}}
\\[0.02\textheight]
{\Large Denys \textsc{Dutykh}}\\
{\it\textcolor{gray}{CNRS--LAMA, Universit\'e Savoie Mont Blanc, France}}
\\[0.02\textheight]
{\Large Dimitrios \textsc{Mitsotakis}}\\
{\it\textcolor{gray}{Victoria University of Wellington, New Zealand}}
\\[0.08\textheight]

\vspace*{1.1cm}

\colorbox{Lightblue}{
  \parbox[t]{1.0\textwidth}{
    \centering\huge\sc
    \vspace*{0.7cm}
    
    \textcolor{bluepigment}{Peregrine's system revisited}
    
    \vspace*{0.7cm}
  }
}

\vfill 

\raggedleft     
{\large \plogo} 
\end{titlepage}


\newpage
\thispagestyle{empty} 
\par\vspace*{\fill}   
\begin{flushright} 
{\textcolor{denimblue}{\textsc{Last modified:}} \today}
\end{flushright}


\newpage
\maketitle
\thispagestyle{empty}


\begin{abstract}

In 1967 D.~H.~\textsc{Peregrine} proposed a \textsc{Boussinesq}-type model for long waves in shallow waters of varying depth \cite{Peregrine1967}. This prominent paper turned a new leaf in coastal hydrodynamics along with contributions by F.~\textsc{Serre} \cite{Serre1953}, A.~E.~\textsc{Green} \& P.~M.~\textsc{Naghdi} \cite{Green1976} and many others since then. Several modern \textsc{Boussinesq}-type systems stem from these pioneering works. In the present work we revise the long wave model traditionally referred to as the \textsc{Peregrine} system. Namely, we propose a modification of the governing equations which is asymptotically similar to the initial model for weakly nonlinear waves, while preserving an additional symmetry of the complete water wave problem. This modification procedure is called the \emph{invariantization}. We show that the improved system has well conditioned dispersive terms in the swash zone, hence allowing for efficient and stable run-up computations.

\bigskip
\noindent \textbf{\keywordsname:} Long dispersive waves; Boussinesq equations; Galilean invariance; wave run-up \\


\end{abstract}


\newpage
\tableofcontents
\thispagestyle{empty}


\newpage
\section{Introduction}

Nowadays, \textsc{Boussinesq}-type equations have become the models of choice in the near-shore hydrodynamics. Proposed for the first time in 1871 by J.~\textsc{Boussinesq} \cite{bouss}, these equations have been substantially improved in works by F.~\textsc{Serre} (1953) \cite{Serre1953}, D.~H.~\textsc{Peregrine} (1967) \cite{Peregrine1967}, A.~E.~\textsc{Green} \& P.~M.~\textsc{Naghdi} (1976) \cite{Green1976} and many others\footnote{The steady version of the celebrated \textsc{Serre}--\textsc{Green}--\textsc{Naghdi} equations can be traced back up to Lord \textsc{Rayleigh} \cite{LordRayleigh1876}.}. Nowadays it is almost impossible to list all the bibliography on this subject. Since several decennaries researchers have essentially focused their effort on extending the validity of these models from shallow waters to intermediate depths \cite{Nwogu1993, Madsen1999, Madsen03} under the increasing demand of the coastal engineering community. We refer to \cite{Brocchini2013} for a recent \emph{reasoned} review of this topic. The derivation of of these equations on flat geometries was reviewed in \cite{Khakimzyanov2016c} and the spherical case was covered in \cite{Khakimzyanov2016a}.

The true success of \textsc{Boussinesq} type equations has to deal with the description of the wave breaking phenomenon. Classical Nonlinear Shallow Water Equations (NSWE) predict waves to break too early. Thus, the validity region of NSWE is limited only to the inner surf zone. The success story of \textsc{Boussinesq} systems begins when they were shown to model fairly well breaking waves (see \cite{Zelt1991}). However, the research on robust and efficient numerical methods lags behind the current state of the art in the modeling \cite{Bellotti2001, EIK, Benkhaldoun2008}. Main problems arise from the numerical treatment of the shoreline and the stability of the resulting method. Most of computational algorithms run into numerical troubles when a sufficiently big amplitude wave reaches the run-up region. These problems are obviously due to the uncontrolled numerical instabilities coming from the dispersive terms discretization (see \cite{Bellotti2001}). These difficulties were reported presumably for the first time in P.~\textsc{Madsen} \etal (1997) \cite{Madsen1997} (this emphasis is ours):
\begin{quote}
``However, to make this technique [slot technique] operational in connection with Boussinesq type models a couple of problems call for special attention. [\,\dots\,] Firstly \emph{the Boussinesq terms are switched off} at the still water shoreline, where their relative importance is extremely small anyway. Hence in this region the equations simplify to the nonlinear shallow water equations.''
\end{quote}
This extremely pragmatic point of view is still shared nowadays by a number of researchers. However, in our opinion, it is the model which has to decide naturally whether the dispersion is important or not. Ideally, the treatment of dry areas today should be as simple and natural as the treatment of shock waves in shock-capturing schemes \cite{Toro2009}. In this study we present a fully dispersive numerical simulation of a wave run-up on a complex beach where dispersive terms are present in the entire domain.

The main idea of this study is to revise the original \textsc{Peregrine} system \cite{Peregrine1967}. Some properties of the complete water wave problem have been lost as a price to pay for the model simplification. Namely, as for many other models derived by asymptotic methods, we loose the invariance under vertical translations. If no special care is taken, we inevitably loose this property, since the asymptotic expansion is performed in a very particular frame of reference (around the mean water level $z\ =\ 0$). However, the full water wave problem possesses this symmetry (\cf~~\cite{Benjamin1982}).

The model we propose in this study is asymptotically similar to the original system since we add only higher order contributions which are formally negligible while greatly improving structural properties of the model. Consequently, the linear dispersion relation of the original system is conserved as well. The great improvement consists in dispersive terms which are better conditioned from the numerical point of view and they fit better our physical intuition about their relative importance when we approach the shoreline. A similar attempt of improving dispersive terms by adding nonlinear contributions was also undertaken recently in \cite{Bellotti2002, Antuono2009}. The procedure presented in this study is sometimes referred to in the literature as the invariantization process. Conservative versions of some \textsc{Nwogu}-type systems have been proposed in \cite{Filippini2015, Bellec2016}.

The present study is organized as follows. In Section~\ref{sec:models} we present some rationale on the \textsc{Peregrine} system and its invariantization, with particular emphasis on the numerical generation of solitary wave solutions of the modified system, which are studied in Section~\ref{sec:sw}. Some elements on the numerical discretization by the finite volume method are given in Section~\ref{sec:fv}. Then, some numerical results are shown in Sections~\ref{sec:nums} while applications to waves generated due to landslides are presented  in~\ref{sec:landslide}. Finally, the main conclusions and perspectives of this study are outlined in Section~\ref{sec:concl}.


\section{Mathematical modeling}
\label{sec:models}

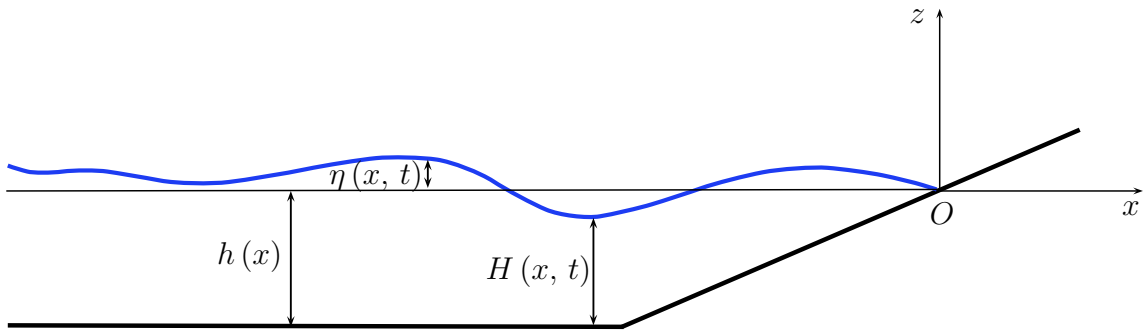
\begin{figure}
\scalebox{0.99} 
{
\begin{pspicture}(0,-2.1967187)(15.319062,2.2067187)
\definecolor{color19}{rgb}{0.11372549019607843,0.23529411764705882,0.9490196078431372}
\psline[linewidth=0.06cm](0.02,-2.1467187)(8.2,-2.1667187)
\psline[linewidth=0.06cm](8.18,-2.1667187)(14.26,0.47328126)
\pscustom[linewidth=0.055999998,linecolor=color19]
{
\newpath
\moveto(12.36,-0.32671875)
\lineto(12.1688175,-0.26671875)
\curveto(12.073225,-0.23671874)(11.808176,-0.17171875)(11.638719,-0.13671875)
\curveto(11.46926,-0.10171875)(11.104275,-0.05171875)(10.908747,-0.03671875)
\curveto(10.713218,-0.02171875)(10.283056,-0.05171875)(10.048423,-0.09671875)
\curveto(9.813789,-0.14171875)(9.409698,-0.24671875)(9.240239,-0.30671874)
\curveto(9.070782,-0.36671874)(8.736211,-0.47671875)(8.571098,-0.52671874)
\curveto(8.405987,-0.57671875)(8.062725,-0.65671873)(7.8845778,-0.68671876)
\curveto(7.7064295,-0.71671873)(7.3501344,-0.6717188)(7.171986,-0.5967187)
\curveto(6.993838,-0.52171874)(6.641887,-0.33171874)(6.468085,-0.21671875)
\curveto(6.2942824,-0.10171875)(5.9206057,0.04328125)(5.720732,0.07328125)
\curveto(5.5208592,0.10328125)(5.1167674,0.11328125)(4.912549,0.09328125)
\curveto(4.7083306,0.07328125)(4.2390633,-0.00171875)(3.9740136,-0.05671875)
\curveto(3.7089648,-0.11171875)(3.2005913,-0.19671875)(2.9572675,-0.22671875)
\curveto(2.7139435,-0.25671875)(2.2577112,-0.24671875)(2.044803,-0.20671874)
\curveto(1.8318945,-0.16671875)(1.4799439,-0.10671875)(1.3409015,-0.08671875)
\curveto(1.2018591,-0.06671875)(0.90639436,-0.06671875)(0.7499719,-0.08671875)
\curveto(0.5935492,-0.10671875)(0.35891542,-0.10671875)(0.28070435,-0.08671875)
\curveto(0.20249298,-0.06671875)(0.09821106,-0.03671875)(0.02,-0.00671875)
}
\psline[linewidth=0.02cm,arrowsize=0.05291667cm 2.0,arrowlength=1.4,arrowinset=0.4]{->}(12.4,-0.34671876)(12.4,2.0932813)
\psline[linewidth=0.02cm](12.4,-0.32671875)(0.0,-0.34671876)
\psline[linewidth=0.02cm,arrowsize=0.05291667cm 2.0,arrowlength=1.4,arrowinset=0.4]{->}(12.38,-0.34671876)(15.1,-0.34671876)
\psline[linewidth=0.025cm,arrowsize=0.05291667cm 2.0,arrowlength=1.4,arrowinset=0.4]{<->}(3.78,-0.34671876)(3.78,-2.1667187)
\usefont{T1}{ptm}{m}{n}
\rput(3.2345312,-1.2217188){$h\,(x)$}
\psline[linewidth=0.025cm,arrowsize=0.05291667cm 2.0,arrowlength=1.4,arrowinset=0.4]{<->}(5.6,-0.30671874)(5.6,0.07328125)
\usefont{T1}{ptm}{m}{n}
\rput(4.914531,-0.14171875){$\eta\,(x,\,t)$}
\psline[linewidth=0.025cm,arrowsize=0.05291667cm 2.0,arrowlength=1.4,arrowinset=0.4]{<->}(7.8,-0.70671874)(7.8,-2.1467187)
\usefont{T1}{ptm}{m}{n}
\rput(7.074531,-1.4217187){$H\,(x,\,t)$}
\usefont{T1}{ptm}{m}{n}
\rput(12.434531,-0.64171875){$O$}
\usefont{T1}{ptm}{m}{n}
\rput(14.944531,-0.58171874){$x$}
\usefont{T1}{ptm}{m}{n}
\rput(12.104531,1.9982812){$z$}
\end{pspicture}
}
\caption{\small\em Sketch of the fluid domain with a sloping beach.}
\label{fig:sketch}
\end{figure}

Consider a \textsc{Cartesian} coordinate system in two space dimensions $(x,\,z)$ to simplify the notation. The $z-$axis is taken vertically upwards and the $x-$axis is horizontal and coincides traditionally with the still water level. The fluid domain is bounded below by the bottom $z\ =\ -h\,(x)$ and above by the free surface $z\ =\ \eta\,(x,\,t)\,$. Below we will also need the total water depth $H\,(x,\,t)\ \eqdef\ h\,(x)\ +\ \eta\,(x,\,t)\,$. The sketch of the fluid domain is given in Figure~\ref{fig:sketch}. The flow is supposed to be incompressible and the fluid is inviscid. An additional simplifying assumption of the flow irrotationality is traditionally made as well.

\begin{remark}
We would like to underline the fact that in the presence of a free surface the vorticity does not remain zero even if it is so initially. A singularity at the free surface (\eg the wave breaking) may lead to vortex sheets creation. However, the water wave theory is not supposed to hold when a wave breaking event occurs.
\end{remark}

Under the previously described physical assumptions, D.~H.~\textsc{Peregrine} (1967) \cite{Peregrine1967} derived the following system of equations which is valid in the Boussinesq long wave regime:
\begin{equation}\label{eq:p1}
  \eta_{\,t}\ +\ \bigl((h\ +\ \eta)\,u\bigr)_{\,x}\ =\ 0\,,
\end{equation}
\begin{equation}\label{eq:p2}
  u_{\,t}\ +\ u\,u_{\,x}\ +\ g\,\eta_{\,x}\ -\ \frac{h}{2}\, (h\,u)_{\,x\,x\,t}\ +\ \frac{h^2}{6}\;u_{\,x\,x\,t}\ =\ 0\,,
\end{equation}
where $u\,(x,\,t)$ is the depth averaged fluid velocity, $g$ is the gravity acceleration and under-scripts ($u_{\,x} \eqdef \pd{u}{x}\,$, $\eta_{\,t} \eqdef \pd{\eta}{t}$) denote partial derivatives.


\subsection{Symmetry analysis}
\label{sec:sym}

In this Section we assume the bottom to be flat, \ie~$h\ =\ d\ =\ \const\ >\ 0\,$. Otherwise, bathymetry variations will destroy a part of symmetries of the governing equations \eqref{eq:p1}, \eqref{eq:p2}. The infinitesimal generators of symmetries transformations for the classical \textsc{Peregrine} system are given here:
\begin{align*}
  \X_{\,1}\ &=\ \pd{}{t}\,, \\
  \X_{\,2}\ &=\ \pd{}{x}\,, \\
  \X_{\,3}\ &=\ u\;\pd{}{u}\ +\ 2\,(\eta\ +\ d)\;\pd{}{\eta}\ -\ t\;\pd{}{t}\,. \\
\end{align*}
It is not difficult to see that the generator $\X_{\,1}$ corresponds to time translations:
\begin{equation*}
  \tilde{t}\ =\ t\ +\ \eps_{\,1}\,, \qquad
  \tilde{x}\ =\ x\,, \qquad
  \tilde{\eta}\ =\ \eta\,, \qquad
  \tilde{u}\ =\ u\,.
\end{equation*}
Similarly, the generator $\X_{\,2}$ gives translations in space:
\begin{equation*}
  \tilde{t}\ =\ t\,, \qquad
  \tilde{x}\ =\ x\ +\ \eps_{\,2}\,, \qquad
  \tilde{\eta}\ =\ \eta\,, \qquad
  \tilde{u}\ =\ u\,.
\end{equation*}
Finally, the generator $\X_{\,3}$ is nothing else but a scaling transformation:
\begin{equation*}
  \tilde{t}\ =\ \ue^{-\eps_{\,3}}\,t\,, \qquad
  \tilde{x}\ =\ x\,, \qquad
  \tilde{\eta}\ =\ \ue^{2\,\eps_{\,3}}\,(d\ +\ \eta)\ -\ d\,, \qquad
  \tilde{u}\ =\ \ue^{\eps_{\,3}}\,u\,.
\end{equation*}
There are no other symmetry transformations of the classical \textsc{Peregrine} system. If this system possessed a \textsc{Lagrangian} structure, we could employ \textsc{Noether} theorem to convert symmetries to conservation laws \cite{Olver1993}. For instance, space translations $\X_{\,2}$ correspond to the momentum conservation. The time translations $\X_{\,1}$ would yield the energy conservation equation, if we only could apply the \textsc{Noether} theorem. This is one of the reasons why it is widely believed that the classical \textsc{Peregrine} system has no energy functional. However, using some other complementary methods \cite{Bluman2010, Cheviakov2010a} we were able to compute an additional conservation law, which can be associated to the energy:
\begin{multline*}
  \Bigl(\half\;u^{\,2}\ +\ g\,(d\ +\ \eta)\,\ln(d\ +\ \eta)\ -\ g\,\eta\ -\ \frac{d^{\,2}}{6}\;u\,u_{\,x\,x}\Bigr)_{\,t}\ +\\
  \Bigl[\,\third\;u^{\,3}\ +\ g\,u\,(d\ +\ \eta)\,\ln(d\ +\ \eta)\ +\ \frac{d^{\,2}}{6}\;u_{\,x}\,u_{\,t}\ -\ \frac{d^{\,2}}{6}\;u\,u_{\,t\,x}\,\Bigr]_{\,x}\ =\ 0\,.
\end{multline*}
The last conservation law can be used, for example, to check the accuracy of numerical schemes over even bottoms for the sake of validation. In some situations additional conservation laws might be used in theoretical investigations as well.


\subsection{Dimensionless equations}

Some of our developments below will be more transparent if we work in dimensionless variables. The classical long wave scaling is the following:
\begin{equation*}
  x^{\,\prime} \eqdef \frac{x}{\ell}\,, \quad z^{\,\prime} \eqdef \frac{z}{h_{\,0}}\,, \quad t^{\,\prime} \eqdef \frac{g}{h_{\,0}}\;t\,, \quad
  \eta^{\,\prime} \eqdef \frac{\eta}{a}\,, \quad u^{\,\prime} \eqdef \frac{u}{\sqrt{g\,h_{\,0}}}\,,
\end{equation*}
where $h_{\,0}\,$, $a\,$, $\ell$ are the characteristic water depth, wave amplitude and wave length respectively. The following dimensionless numbers are defined from them:
\begin{equation*}
  \eps \eqdef \frac{a}{h_{\,0}}\,, \quad \mu^{\,2} \eqdef \Bigl(\frac{h_{\,0}}{\ell}\Bigr)^{\,2}\,, \quad
  \St \eqdef \frac{\eps}{\mu^{\,2}}\,.
\end{equation*}
Parameters $\eps$ and $\mu^{\,2}$ measure the wave nonlinearity and dispersion, while the so-called \textsc{Stokes}--\textsc{Ursell} number $\St$ measures the relative importance of these effects. In the \textsc{Boussinesq} regime the \textsc{Stokes}--\textsc{Ursell} number is supposed to be of the order of one, \ie $\St\ \sim\ 1\,$. The importance of this parameter is discussed by \eg~F.~\textsc{Ursell} (1953) \cite{Ursell1953}. The \textsc{Peregrine} system \eqref{eq:p1}, \eqref{eq:p2} in scaled variables at the order $\O\,(\eps\ +\ \mu^{\,2})$ reads (primes are dropped below for the sake of convenience):
\begin{equation*}
  \eta_{\,t}\ +\ \bigl((h\ +\ \eps\,\eta)\,u\bigr)_{\,x}\ =\ 0\,,
\end{equation*}
\begin{equation*}
  u_{\,t}\ +\ \eps\, u\,u_{\,x}\ +\ \eta_{\,x}\ -\ \mu^{\,2}\,\Bigl(\frac{h}{2}\;(h\,u)_{\,x\,x\,t}\ -\ \frac{h^{\,2}}{6}\; u_{\,x\,x\,t}\Bigr)\ =\ \O(\eps^{\,2}\ +\ \eps\,\mu^{\,2}\ +\ \mu^{\,4})\,,
\end{equation*}
where on the right hand side of the last equation we put the order of neglected terms. Since the \textsc{Stokes}--\textsc{Ursell} number $\St\ \sim\ 1\,$, we have asymptotic similarity relations in the \textsc{Boussinesq} regime:
\begin{equation*}
  \eps^{\,2}\ \sim\ \eps\,\mu^{\,2}\ \sim\ \mu^{\,4}\,.
\end{equation*}


\subsection{Vertical translations}

In this section we examine an important property of the water wave problem --- invariance under vertical translations (subgroup $G_{\,5}$ in Theorem~4.2, T.~\textsc{Benjamin} \& P.~\textsc{Olver} (1982) \cite{Benjamin1982}). This transformation is described by the following simple change of variables:
\begin{equation}\label{eq:vert}
  z\ \leftarrow\ z\ +\ d\,, \quad \eta\ \leftarrow\ \eta\ -\ d\,, \quad
  h\ \leftarrow\ h\ +\ d\,, \quad u\ \leftarrow\ u\,,
\end{equation}
where $d$ is some constant. Here again, it is straightforward to check that the mass conservation Equation~\eqref{eq:p1} remains invariant under transformation \eqref{eq:vert}, while Equation~\eqref{eq:p2} produces many additional dispersive terms proportional to the constant translation $d\,$:
\begin{equation*}
  u_{\,t}\ +\ u\,u_{\,x}\ +\ g\,\eta_{\,x}\ -\ \frac{h}{2}\;(h\,u)_{\,x\,x\,t}\ +\ \frac{h^{\,2}}{6}\;u_{\,x\,x\,t}\ -\ \frac{h\,d}{6}\;u_{\,x\,x\,t}\ -\ \frac{d^{\,2}}{3}\,u_{\,x\,x\,t}\ -\ \frac{d}{2}\;(h\,u)_{\,x\,x\,t}\ =\ 0\,.
\end{equation*}
The reason for this discrepancy is that the coefficient $h^{\,n}\,(x)$ $(n\ =\ 1,\,2)$ in front of the dispersive terms is not invariant under the vertical shift. The right variable to use is the total water depth $H\,(x,\,t)\ =\ h\,(x)\ +\ \eta\,(x,\,t)$ which is independent of the chosen coordinate reference frame. Here again, the discrepancy is a result of the asymptotic expansion around the still water level. Consequently, the derived model is valid only for this particular choice of the coordinate axis $O\,x\,$. To make System \eqref{eq:p1}, \eqref{eq:p2} frame independent we shall add higher order nonlinear terms which are asymptotically negligible but have important implications in structural properties of the resulting model.

In dimensionless variables the total water depth is expressed as $H\,(x,\,t)\ =\ h\,(x)\ +\ \eps\,\eta\,(x,\,t)\,$. As a corollary, we obtain two asymptotic relations which will be used below:
\begin{equation*}
  h\ =\ H\ +\ \O(\eps)\,, \qquad h_{\,x}\ =\ H_{\,x}\ +\ \O(\eps)\,, \qquad H_{\,t}\ =\ \O(\eps)\,.
\end{equation*}
Mathematically it means that the bathymetry function should be completed by an $\O(\eps)$ term to become invariant under vertical translations. While performing this invariantization, we will also recast our model in conservative variables $(H,\, Q)\,$, where $Q \eqdef H\,u$ is the horizontal momentum. This modification will allow us to employ those numerical methods developed in the literature for the discretization of Nonlinear Shallow Water Equations (NSWE) \cite{Zhou2002, DeKaKa, Dutykh2009a, Dutykh2010}.

The mass conservation Equation~\eqref{eq:p1} in the new variables trivially reads:
\begin{equation}\label{eq:cons1}
  H_{\,t}\ +\ Q_{\,x}\ =\ 0\,,
\end{equation}
while the momentum conservation Equation~\eqref{eq:p2} will require more computations. First of all, we multiply Equation~\eqref{eq:cons1} by $u\,$, Equation~\eqref{eq:p2} by $H$ and add them to have:
\begin{equation}\label{eq:pp}
  (Hu)_t\ +\ \bigl(\eps\,H\,u^{\,2}\ +\ \frac{1}{2\,\eps}\;H^{\,2}\bigr)_{\,x}\ -\ \mu^{\,2}\Bigl(\frac{H\,h}{2}\;(h\,u)_{\,x\,x\,t}\ -\ \underbrace{\frac{H\,h^{\,2}}{6}\;u_{\,x\,x\,t}}_{(**)}\Bigr)\ =\ \frac{1}{\eps}\;H\,h_{\,x}\,.
\end{equation}
In the perspective of writing governing equations in the conservative form, the term (**) has to be transformed using this relation:
\begin{equation*}
  u_{\,x\,x}\ \equiv\ \Bigl(\frac{h\,u}{h}\,\Bigr)_{\,x\,x}\ =\ (h\,u)\Bigl(2\;\frac{h_{\,x}^{\,2}}{h^{\,3}}\ -\ \frac{h_{\,x\,x}}{h^{\,2}}\Bigr)\ -\ 2\,\frac{h_{\,x}}{h^{\,2}}\;(h\,u)_{\,x}\ +\ \frac{1}{h}\;(h\,u)_{\,x\,x}\,.
\end{equation*}
Consequently, after simple computations, Equation~\eqref{eq:pp} takes the form:
\begin{multline*}
  (H\,u)_{\,t}\ +\ \bigl(\eps\, H\,u^{\,2}\ +\ \frac{1}{2\,\eps}\;H^{\,2}\bigr)_{\,x}\ -\ \mu^{\,2}\Bigl(\frac{H\,h}{3}\;(h\,u)_{\,x\,x\,t}\ +\ \frac{H}{3}\;(h\,u)(h\,u)_{\,x\,x\,x} \\ +\ \frac{H\,h_{\,x}}{3}\;(h\,u)_{\,x\,t}\ -\ \frac13\;\bigl(\frac{H}{h}\;h_{\,x}^{\,2}\ -\ \frac12\;H\,h_{\,x\,x}\bigr)\,(h\,u)_{\,t}\Bigr)\ =\ \frac{1}{\eps}\;H\,h_{\,x}\,.
\end{multline*}
The last equation is ready for the invariantization process. For illustrative purposes we show these computations only for the first dispersive term:
\begin{equation*}
  \mu^{\,2}\;\frac{H\,h}{3}\;(h\,u)_{\,x\,x\,t}\ =\ \frac{\mu^{\,2}}{3}\; H^{\,2}\,(H\,u)_{\,x\,x\,t}\ +\ \O(\eps\,\mu^{\,2})\ =\ \frac{\mu^{\,2}}{3}\;H^{\,2}\,Q_{\,x\,x\,t}\ +\ \O(\eps\,\mu^{\,2})\,.
\end{equation*}
Thus we add again only higher order terms which have no impact onto linear dispersive characteristics of the initial system. By proceeding in an analogous manner with all other dispersive terms and turning back to dimensional variables we obtain the following momentum conservation equation:
\begin{multline}\label{eq:cons2}
  \Bigl(1\ +\ \frac13\; H_{\,x}^{\,2}\ -\ \frac16\; H\,H_{\,x\,x}\Bigr)\,Q_{\,t}\ -\ \frac13\; H^{\,2}\,Q_{\,x\,x\,t}\ -\ \frac13\; H\,H_{\,x}\,Q_{\,x\,t}\\
   +\ \Bigl(\frac{Q^{\,2}}{H}\ +\ \frac{g}{2}\,H^{\,2}\Bigr)_{\,x}\ =\ g\,H\, h_{\,x}\,.
\end{multline}

The system \eqref{eq:cons1}, \eqref{eq:cons2} (that will be called the modified \textsc{Peregrine} system or, in a short-hand notation, the m-\textsc{Peregrine} system) actually has more advantages than being simply invariant under two additional transformations. The added value of this invariantization process goes far beyond the initial symmetry consideration. Namely, in this way we extend the system validity to the run-up process and improve numerical conditioning of dispersive terms. For the first time Equations \eqref{eq:cons1}, \eqref{eq:cons2} were used and validated for wave run-up problems in \cite{Dutykh2011e}.

In natural environments, dispersive effects become gradually less and less important when a wave travels shore-ward to become negligible in the shoreline vicinity. This is the reason why NSWE can be successfully used to describe to some extent the run-up process. This physical observation can be translated into the mathematical language by the condition that dispersive terms go to zero when the total water depth vanishes. If this condition is not fulfilled, numerical instabilities may appear as reported by G.~\textsc{Bellotti} \& M.~\textsc{Brocchini} (2002) \cite{Bellotti2002}:
\begin{quote}
  ``In our attempt to use these equations from intermediate waters up to the shoreline (see Bellotti and Brocchini, 2001) we run into numerical troubles when reaching the run-up region, i.e. $x\ >\ 0\,$. These problems were essentially related to numerical instabilities due to the uncontrolled growth of the dispersive contributions (i.e. $\O(\mu^{\,2})$-terms).''
\end{quote}

The reason for the extended numerical stability of the proposed model is twofold. First of all, in the numerical algorithm we have to invert at some point an elliptic operator written over the time derivative in Equation~\eqref{eq:cons2}:
\begin{equation*}
\Bigl(1\ +\ \frac13\; H_{\,x}^{\,2}\ -\ \frac16\; H\,H_{\,x\,x}\Bigr)\,q\ -\ \frac13\; H^{\,2}\,q_{\,x\,x}\ -\ \frac13\; H\,H_{\,x}\,q_{\,x}\ =\ W\,,
\end{equation*}
where $W$ is a known function arising from the advective terms discretization. It turns out that the resulting linear system is better conditioned if the model is written in terms of the total water depth. The second stability advantage comes from the fact that almost all dispersive terms naturally vanish as we approach the shoreline.

\begin{remark}
It is noted that the same invariantization technique can be used also for the case of moving bottom bathymetry and for higher dimensions, \cf~Section \ref{sec:landslide}.
\end{remark}


\subsubsection{Symmetry analysis}

The symmetries of the m-\textsc{Peregrine} system (over flat bottom) can be computed using the standard methods as we did for the classical counterpart in Section~\ref{sec:sym}. The dimension of the symmetry group turns out to be the same as above. The infinitesimal generators are given below:
\begin{align*}
  \X_{\,1}\ &=\ \pd{}{t}\,, \\
  \X_{\,2}\ &=\ \pd{}{x}\,, \\
  \X_{\,3}\ &=\ t\;\pd{}{t}\ +\ 2\,x\;\pd{}{x}\ +\ 2\,H\;\pd{}{H}\ +\ 3\,Q\;\pd{}{Q}\,.\\
\end{align*}
The generated symmetry transformations are essentially the same. Generator $\X_{\,1}$ yields time translations:
\begin{equation*}
  \tilde{t}\ =\ t\ +\ \eps_{\,1}\,, \qquad
  \tilde{x}\ =\ x\,, \qquad
  \tilde{H}\ =\ H\,, \qquad
  \tilde{Q}\ =\ Q\,,
\end{equation*}
while $\X_{\,2}$ gives translations in space:
\begin{equation*}
  \tilde{t}\ =\ t\,, \qquad
  \tilde{x}\ =\ x\ +\ \eps_{\,2}\,, \qquad
  \tilde{H}\ =\ H\,, \qquad
  \tilde{Q}\ =\ Q\,.
\end{equation*}
Finally, $\X_{\,3}$ is a scaling transformation\footnote{Notice, please, that this scaling is different from $\X_{\,3}$ given in Section~\ref{sec:sym}.}:
\begin{equation*}
  \tilde{t}\ =\ \ue^{\,\eps_{\,3}}\,t\,, \qquad
  \tilde{x}\ =\ \ue^{\,2\,\eps_{\,3}}\,x\,, \qquad
  \tilde{H}\ =\ \ue^{\,2\,\eps_{\,3}}\,H\,, \qquad
  \tilde{Q}\ =\ \ue^{\,3\,\eps_{\,3}}\,Q\,.
\end{equation*}


\subsubsection{Pressure distribution}

For some practical applications we need to estimate the pressure field inside the fluid and more particularly at the bottom. For example, the operational NOAA Tsunami Warning System heavily relies on a network of DART buoys detecting tsunami waves by measuring the pressure at the ocean bottom \cite{Titov2005, Bernard2007}. In this section we propose a way to reconstruct the pressure field in the whole water column.

In the original work of D.~H.~\textsc{Peregrine} \cite{Peregrine1967} one can find the following correct asymptotic expansion for the pressure field:
\begin{equation}\label{eq:PerPress}
  p\ =\ -z\ +\ \eps\,\eta\ +\ \mu^{\,2}\,\bigl(z\,(h\,u)_{\,x\,t}\ +\ \half\, z^{\,2}\, u_{\,x\,t}\bigr)\ +\ \O(\eps^{\,2}\ +\ \eps\,\mu^{\,2}\ +\ \mu^{\,4})\,.
\end{equation}
The first two terms on the right hand side correspond to the usual hydrostatic pressure while the last two terms are purely non-hydrostatic contributions brought by dispersive effects.

However, the original expression \eqref{eq:PerPress} for the pressure given by the asymptotic expansion method has one important drawback. Namely, it satisfies the free surface dynamic boundary condition $\left.p\right\vert_{\,z\,=\,\eps\,\eta}\ =\ 0$ only to the leading order. Consequently, the first improvement we propose is to add some specific higher order terms to recover this property at all orders we retain in the equation:
\begin{equation*}
  p\ \approx\ -z\ +\ \eps\,\eta\ +\ \mu^{\,2}\, \bigl((z\ -\ \eps\,\eta)\,(h\,u)_{\,x\,t}\ +\ \half\;\mu^{\,2}\,(z\ -\ \eps\,\eta)^{\,2}\,u_{\,x\,t}\bigr)\,.
\end{equation*}
Now we will make a transformation consistent with the modified \textsc{Peregrine} system \eqref{eq:cons1}, \eqref{eq:cons2} which consists in replacing $h$ by its asymptotically equivalent \emph{and} invariant by vertical translations counterpart $H$ in the third term\footnote{The asymptotic argument holds here since this term is $\O(\mu^{\,2})\,$.} of the last formula:
\begin{equation*}
  p\ \approx\ -z\ +\ \eps\,\eta\ +\ \mu^{\,2} \bigl((z\ -\ \eps\,\eta)\,(H\,u)_{\,x\,t}\ +\ \half\;\mu^{\,2}\,(z\ -\ \eps\,\eta)^{\,2} u_{\,x\,t} \bigr)\,.
\end{equation*}
Finally, if we turn back to the dimensional and conservative variables, the final expression for the pressure will take this form:
\begin{equation*}
  \frac{p}{\rho}\ =\ g\,(\eta\ -\ z)\ +\ (z\ -\ \eta)\,Q_{\,x\,t}\ +\ \frac12\;(z\ -\ \eta)^{\,2}\Bigl(\frac{Q}{H}\Bigr)_{\,x\,t}\,.
\end{equation*}
where $\rho$ is the constant fluid density. It is straightforward now to compute the pressure value at the bottom by evaluating the last expression at $z\ =\ -h\,$:
\begin{equation*}
  \left.\frac{p}{\rho}\right\vert_{\,z\,=\,-h}\ =\ g\,H\ +\ H\,Q_{\,x\,t}\ +\ \frac12\;H^{\,2}\,\Bigl(\frac{Q}{H}\Bigr)_{\,x\,t}\,.
\end{equation*}
The latter can be directly used, for example, to compute synthetic pressure records which can be compared with real observations in deep ocean \cite{Titov2005}.


\subsection{Galilean invariance}

In the same line of ideas, there is a question of the \textsc{Galilean} invariance of various \textsc{Boussinesq}-type equations. In this section we check whether the \textsc{Peregrine} system \eqref{eq:p1}, \eqref{eq:p2} remains invariant under the \textsc{Galilean} transformation. This issue was already addressed in the context of some other systems by C.~I.~\textsc{Christov} (2001) \cite{Christov2001}.

The procedure is classical. First of all, we assume throughout this section the bottom to be flat $h\ =\ \const$. We choose another frame of reference which moves uniformly rightwards with constant celerity $c\,$. Analytically it is expressed by the following change of variables:
\begin{equation}\label{eq:boost}
  x\ \leftarrow\ x\ -\ c\,t\,, \quad t\ \leftarrow\ t\,, \quad \eta\,(x,\,t)\ \leftarrow\ \eta\,(x\,-\,c\,t,\,t)\,, \quad u\,(x,\,t)\ \leftarrow\ u\,(x\,-\,c\,t,\,t)\ +\ c\,,
\end{equation}
After some simple computations, one can easily check that the mass conservation Equation~\eqref{eq:p1} remains invariant under the \textsc{Galilean} boost \eqref{eq:boost}, while Equation~\eqref{eq:p2} has an extra term (*):
\begin{equation*}
  u_{\,t}\ +\ u\,u_{\,x}\ +\ g\,\eta_{\,x}\ -\ \frac{h}{2}\; (h\,u)_{\,x\,x\,t}\ +\ \frac{h^{\,2}}{6}\;u_{\,x\,x\,t}\ +\ \underbrace{\frac{c\,h^{\,2}}{3}\;u_{\,x\,x\,x}}_{(*)}\ =\ 0\,.
\end{equation*}
Consequently, the \textsc{Peregrine} system in its original form does not possess the very basic \textsc{Galilean} invariance property while the complete water wave problem does (subgroups $G_{\;7,\,8}$ in three dimensions, see Theorem~4.2, T.~\textsc{Benjamin} \& P.~\textsc{Olver} (1982) \cite{Benjamin1982}). Some consequences of this shortcoming are discussed in \textsc{Christov} (2001) \cite{Christov2001}.

In order to recover the broken symmetry we propose to modify Equation~\eqref{eq:p2} in the following way:
\begin{equation}\label{eq:p3}
  u_{\,t}\ +\ u\,u_{\,x}\ +\ g\,\eta_{\,x}\ -\ \frac{h}{2}\; (h\,u)_{\,x\,x\,t}\ +\ \frac{h^{\,2}}{6}\;u_{\,x\,x\,t}\ -\ \frac{h}{3}\;u\,(h\,u)_{\,x\,x\,x}\ =\ 0\,.
\end{equation}
If we perform the same computations as above, we will see that the modified model \eqref{eq:p1}, \eqref{eq:p3} remains invariant under the \textsc{Galilean} boost \eqref{eq:boost}. In order to understand better this modification, we have to switch to dimensionless variables:
\begin{equation*}
  u_{\,t}\ +\ \eps\,u\,u_{\,x}\ +\ \eta_{\,x}\ -\ \mu^{\,2}\Bigl(\frac{h}{2}\;(h\,u)_{\,x\,x\,t}\ -\ \frac{h^{\,2}}{6}\;u_{\,x\,x\,t}\Bigr)\ -\ \eps\,\mu^{\,2}\,\frac{h}{3}\;u\,(h\,u)_{\,x\,x\,x}\ =\ 0\,.
\end{equation*}
Now it is clear that we add a higher order $\O(\eps\,\mu^{\,2})$ nonlinear dispersive term which normally has to be omitted according to the philosophy of asymptotic methods. However, we prefer to retain it to recover an important physical property of the model --- the \textsc{Galilean} invariance.

\begin{remark}
Since the term $\frac{h}{3}\;u\,(h\,u)_{\,x\,x\,x}$ is a nonlinear dispersive term, it has no effect onto linear dispersion characteristics of the original model. The same remark applies to developments presented below as well.
\end{remark}

Consequently, we are able to add a higher order dispersive term to Equation~\eqref{eq:p2} which makes the system \textsc{Galilean} invariant. The invariantization process in variables $(\eta,\, u)$ is straightforward. However, if we rewrite the modified system in terms of the conservative variables $(H,\, Q)$ we loose again the \textsc{Galilean} invariance property. One of the reasons is that transformation \eqref{eq:boost} is more complex in these variables. For example, the following chain rules apply:
\begin{equation*}
  Q_{\,t}\ \rightarrow\ Q_{\,t}\ -\ c\, Q_{\,x}\ +\ c\,(H_{\,t}\ -\ c\, H_{\,x})\,, \qquad
  Q_{\,x}\ \rightarrow\ Q_{\,x}\ +\ c\, H_{\,x}\,.
\end{equation*}
The invariantization of the modified \textsc{Peregrine} system \eqref{eq:cons1}, \eqref{eq:cons2} under the \textsc{Galilean} symmetry remains an open question. The discussion of the \textsc{Galilean} invariance of a few other nonlinear dispersive wave systems can be found in \cite{Duran2013}.


\section{Solitary waves}
\label{sec:sw}

Dispersive wave equations possess an important class of solutions --- the Solitary Waves (SW) which result from a balance between nonlinear and dispersive effects \cite{Sandee1991, Kalisch2004, DDMM, Chambarel2009}. The comprehension of these solutions allows to assess some properties of the dispersive system under consideration. We note that analytical SW solutions are not known even for the classical \textsc{Peregrine} system \cite{Peregrine1967}. We have not been able to construct closed-form solutions to the m-\textsc{Peregrine} system either. Consequently, we will apply numerical methods which allow to approximate them accurately \cite{Yang2010}.

A travelling wave solution has the following form:
\begin{equation*}
  H\,(x,\,t)\ \equiv\ H\,(X)\,, \qquad Q\,(x,\,t)\ \equiv\ Q\,(X)\,, \qquad X\ \eqdef\ x\ -\ c_{\,s}\,t\,,
\end{equation*}
where $c_{\,s}$ is the wave propagation speed in an inertial frame of reference. After substituting this {\em ansatz} into the governing Equations~\eqref{eq:cons1}, \eqref{eq:cons2}, we obtain the following system of two coupled Ordinary Differential Equations (ODEs):
\begin{equation}\label{eq:sw1}
  -c_{\,s}\,H^{\,\prime}\ +\ Q^{\,\prime}\ =\ 0\,,
\end{equation}
\begin{multline}\label{eq:sw2}
  -c_{\,s}\,\Bigl(1\ +\ \frac13\;(H^{\,\prime})^{\,2}\ -\ \frac16\;H\,H^{\,\prime\prime}\Bigr)\,Q^{\,\prime}\ +\ \frac{c_{\,s}}{3}\; H^{\,2}\,Q^{\,\prime\prime\prime}\\
  +\ \frac{c_{\,s}}{3}\; H\,H^{\,\prime}\,Q^{\,\prime\prime}\ +\ \Bigl(\frac{Q^{\,2}}{H}\ +\ \frac{g}{2}\;H^{\,2}\Bigr)^{\,\prime}\ =\ 0\,,
\end{multline}
where functions $H\,(X)$ and $Q\,(X)$ are assumed to be sufficiently smooth, even and decaying to zero along with all their derivatives as $\abs{X}\ \to\ \infty\,$. Throughout this section we will consider the wave propagation over a flat bottom, \ie $h\ \equiv\ \const$.

The former Equation~\eqref{eq:sw1} can be used to eliminate the variable $Q\,(X)$ from the latter equation. It will be more convenient also to work with the free surface elevation $\eta\,(X)\,$:
\begin{eqnarray}\label{eq:sw3}
  L_{\,0}\,\eta\ &=&\ (gh\ -\ c_{\,s}^{\,2})\,\eta^{\,\prime}\ +\ \frac{c_{\,s}^{\,2}\,h^{\,2}}{3}\,\eta^{\,\prime\prime\prime}\ +\ \Bigl(\frac{c_{\,s}^{\,2}\;\eta^{\,2}}{h\ +\ \eta}\Bigr)^{\,\prime}\ +\ \frac{g}{2}\;(\eta^{\,2})^{\,\prime}\ -\ \frac{c_{\,s}^{\,2}}{3}\;(\eta^{\,\prime})^{\,3}\nonumber\\
  &&+\ \frac{c_{\,s}^{\,2}}{3}\;(2\,h\,\eta\ +\ \eta^{\,2})\,\eta^{\,\prime\prime\prime}\ +\ \frac{c_{\,s}^{\,2}}{2}\;(h\ +\ \eta)\eta^{\,\prime}\eta^{\,\prime\prime}\ =\ 0\,.
\end{eqnarray}
Once the free surface elevation $\eta\,(X)$ is determined, the velocity can be found from the mass conservation \eqref{eq:sw1}:
\begin{eqnarray}\label{eq:sw4}
  u\,(X)\ =\ \frac{c_{\,s}\,\eta\,(X)}{h\ +\ \eta\,(X)}\,, \qquad Q\,(X)\ =\ c_{\,s}\,\eta\,(X)\,.
\end{eqnarray}
Solitary wave profiles $(\eta\,(X),\, u\,(X))$ can be obtained numerically by approximating solutions to the differential Equation~\eqref{eq:sw3} and then using \eqref{eq:sw4} to compute the velocity profile.

Several strategies to this end exist in the literature (see \cite{Yang2010} and references therein). The one considered here consists of two steps. First, the \textsc{Newton} method is applied to \eqref{eq:sw3}: from an initial iteration $\eta^{\,[0]}\,(X)$ and if the approximation $\eta^{\,[\nu]}\,(X)\,$, $\nu\ =\ 0,\,1,\,\ldots$ to the profile $\eta\,(X)$ at the $\nu$\up{th} iteration is known, then $\eta^{\,[\nu\,+\,1]}\,(X)$ is obtained by solving the equation
\begin{equation}\label{eq:sw5}
  L^{\,[\nu]}\Delta\eta^{\,[\nu]}\ =\ -L_{\,0}\,\eta^{\,[\nu]}\,,
\end{equation}
where $\Delta \eta^{\,[\nu]}\ \eqdef\ \eta^{\,[\nu\,+\,1]}\ -\ \eta^{\,[\nu]}\,$, $L_{\,0}$ is given by \eqref{eq:sw3} and $L^{\,[\nu]}$ is the linearized operator of Equation~\eqref{eq:sw3} evaluated at $\eta^{\,[\nu]}\,(X)\,$.

The second step of our numerical procedure is the discretization of \eqref{eq:sw5}, which will be inspired by several works of J.~\textsc{Boyd} (for more details see \cite{Boyd1986, Boyd2000, Boyd2002a, Boyd2002b}). For $N\ \geq\ 1$ and large $L\ >\ 0\,$, the system \eqref{eq:sw5} is discretized on the interval $\bigl(-L,\,L\bigr)$ by the collocation points
\begin{equation}\label{eq:sw8}
  x_{\,k}\ =\ -L\ +\ (2k\ +\ 1)\,h\,, \qquad h\ =\ \frac{L}{N}\,, \qquad k\ =\ 0,\,\ldots,\,N\,-\,1\,.
\end{equation}
For $\nu\ =\ 0,\,1,\,\ldots\,$, the approximation $\eta_{\,h}^{\,[\nu]}$ to the $\nu$\up{th} iteration $\eta^{\,[\nu]}$ is sought in the space $\Sp_{\,h}\,$, based on \eqref{eq:sw8}, of trigonometric interpolation polynomials of the form
\begin{equation*}
  Z_{\,h}\,(x)\ =\ \sum_{j\,=\,0}^{N\,-\,1} Z_{\,j}\cos\left(\frac{\pi}{2\,L}\;j\,(x\,+\,L)\right)\,.
\end{equation*}
The discrete version of \eqref{eq:sw5} is then as follows. If $\eta^{\,[\nu]}\ \in\ \Sp_{\,h}$ is known, we search for the incremental term $\Delta \eta_{\,h}^{\,[\nu]}\ \eqdef\ \eta_{\,h}^{\,[\nu\,+\,1]}\ -\ \eta_{\,h}^{\,[\nu]}$ in $\Sp_{\,h}\,$, \ie
\begin{equation*}
  \Delta\eta^{\,[\nu]}\,(x)\ =\ \sum_{j\,=\,0}^{N\,-\,1}\,\alpha_{\,j}^{\,[\nu]}\,\cos\left(\frac{\pi}{2\,L}\;j\,(x\,+\,L)\right)\,,
\end{equation*}
and evaluate \eqref{eq:sw5} at the collocation points \eqref{eq:sw8}. This leads to a linear system for the coefficients $\alpha^{\,[\nu]}\ =\ (\alpha_{\,0}^{\,[\nu]},\,\ldots,\,\alpha_{\,N-1}^{\,[\nu]})^{\top}$ of the form
\begin{equation}\label{eq:sw11}
  L_{\,h}^{\,[\nu]}\,\alpha^{\,[\nu]}\ =\ f^{\,[\nu]}\,,
\end{equation}
where the matrix $L_{\,h}^{\,[\nu]}\ =\ \bigl(L_{\,ij}^{\,[\nu]}\bigr)_{\,i,\,j\,=\,0}^{N\,-\,1}$ and the vector $f^{\,[\nu]}\ =\ (f_{\,0}^{\,[\nu]}\,,\,\ldots,\,f_{\,N\,-\,1}^{\,[\nu]})^{\,\top}$ are computed as:
\begin{equation*}
  L_{\,i\,j}^{\,[\nu]}\ =\ L^{\,[\nu]}\,\cos\left(\frac{\pi}{2\,L}\;j\,(x\ +\ L)\right)\Big\vert_{\,x\,=\,x_{\,i}}\,, \qquad f_{\,k}^{\,[\nu]}\ =\ -L_{\,0}\,\eta^{\,[\nu]}\Big\vert_{\,X\,=\,x_{\,k}}\,,
\end{equation*}
for $i,\,k\ =\ 0,\,\ldots,\,N\,-\,1\,$. We note that the construction of coefficients $f^{\,[\nu]}$ in \eqref{eq:sw11} requires the computation of derivatives of $\eta_{\,h}^{\,[\nu]}$ up to the third order at the points \eqref{eq:sw8}. Finally, in order to pass to the next iteration $\eta_{\,h}^{\,[\nu+1}\,$. Equation~\eqref{eq:sw11} has to be solved. The ill-conditioning of the resulting system is treated using the pseudo-inverse technique combined with the iterative refinement (see \cite{Golub1996, Demmel1997, Higham2002, Boyd2002a} for more details). This method solves Equations~\eqref{eq:sw11} in the least squares sense and the solution has a minimum norm.

The overall iterative process is controlled, in a standard way, by two parameters: (\textit{i}) a maximum number of iterations and (\textit{ii}) a tolerance governing the relative error between two consecutive iterations or the residual error:
\begin{equation}\label{eq:sw12}
  \eps_{\,1}\,[\nu]\ =\ \frac{\norm{\eta^{\,[\nu]}\ -\ \eta^{\,[\nu\,-\,1]}}}{\norm{\eta^{\,[\nu]}}}\,, \qquad \eps_{\,2}\,[\nu]\ =\ \norm{L_{\,0}\,\eta^{\,[\nu]}}\,,
\end{equation}
measured in some norm $\norm{\cdot}$ (in the experiments reported below, both the \textsc{Euclidean} and the maximum norms ($l_{\,\infty}$) were implemented). Thus, the iteration stops when the maximum number of iterations is attained or when any of the errors \eqref{eq:sw12} is below a prescribed tolerance.


\subsection{Numerical results}

The described above numerical procedure will be tested and used now to compute several travelling wave solutions to the m-\textsc{Peregrine} Equations~\eqref{eq:cons1}, \eqref{eq:cons2}. For the sake of convenience, we will solve equations in the dimensionless form which is readily obtained by setting dimensional constants $g\ =\ 1$ and $d\ =\ 1\,$. The tolerance parameter in the control of the iterations is chosen to be equal to $10^{\,-13}\,$. The exact solution to the classical \textsc{Serre} equations \cite{Serre1953, Clamond2009, Dutykh2011a} is chosen as the initial approximation at the first iteration.

\begin{figure}
  \centering
  \includegraphics[width=0.85\textwidth]{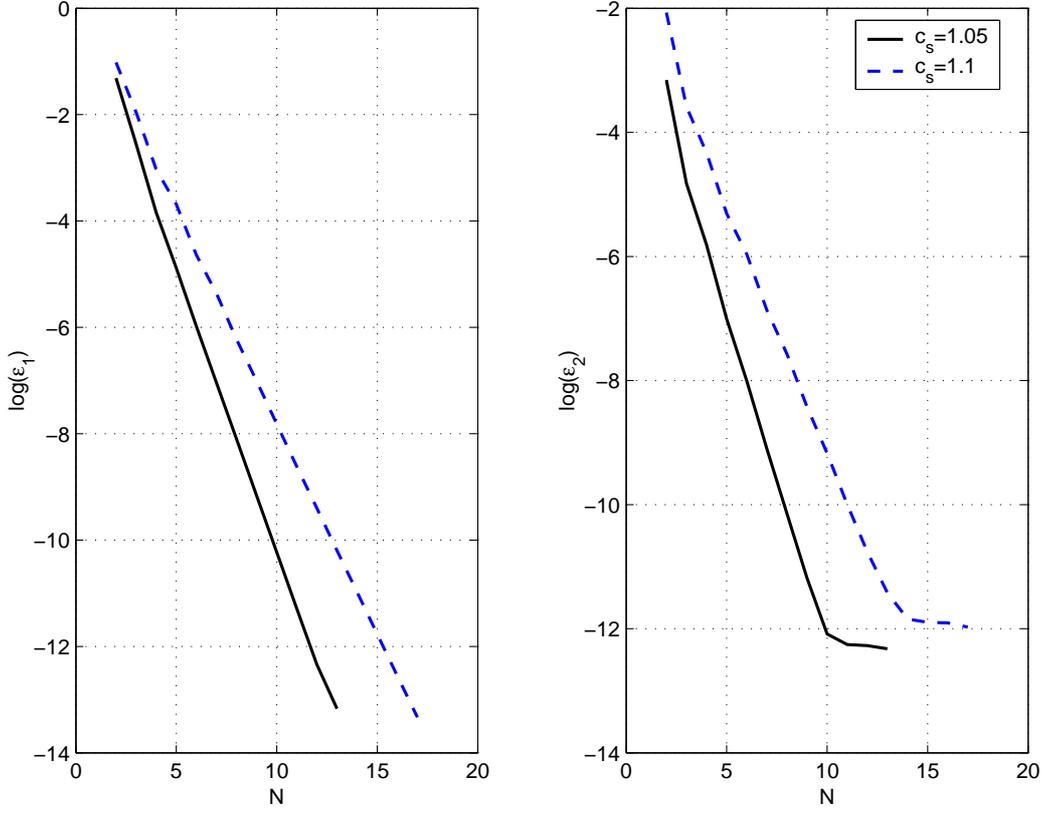}
  \caption{\small\em Decimal logarithm of the relative errors defined in \eqref{eq:sw12}. Relative difference between two iterations $\eps_{\,1}\,[\nu]$ is shown on the left image, while the residual of the equation is depicted on the right. The convergence is illustrated for two values of the propagation velocities $c_{\,s}\ =\ 1.05$ (black solid line) and $1.1$ (blue dashed line).}
  \label{fig:err}
\end{figure}

\begin{figure}
  \centering
  \includegraphics[width=0.85\textwidth]{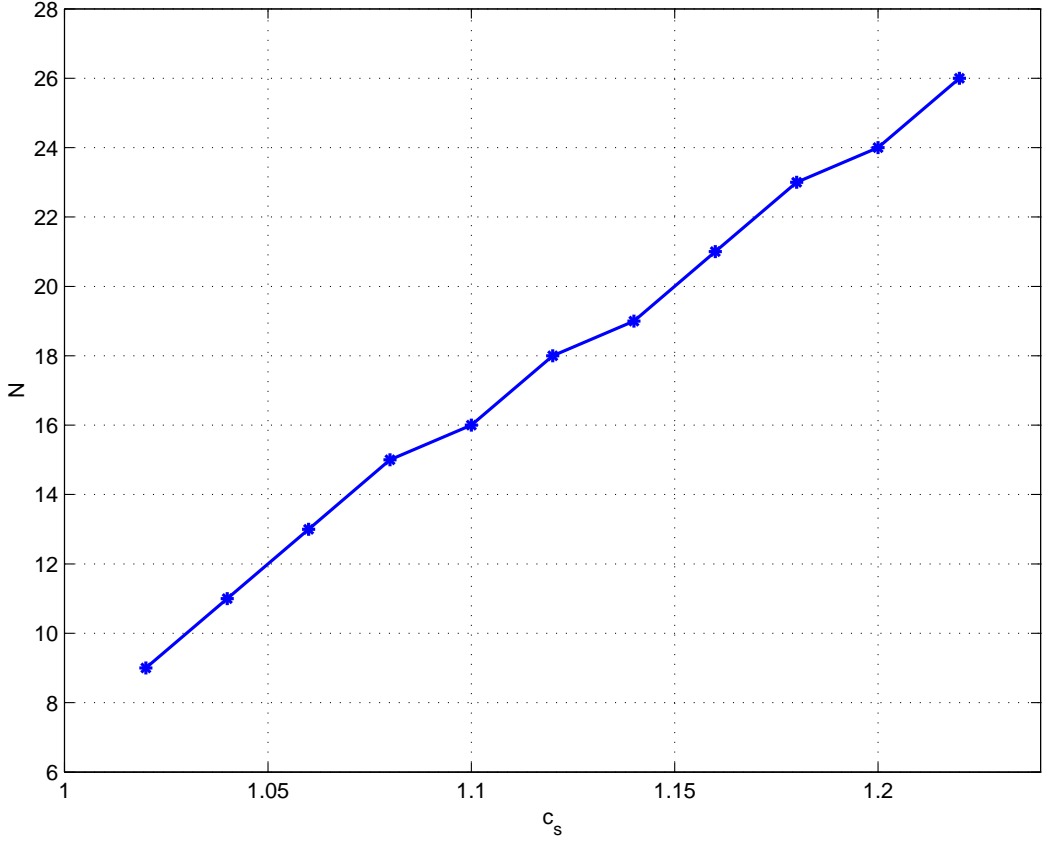}
  \caption{\small\em Dependence of the number of iterations needed to achieve the convergence on the solitary wave propagation speed $c_{\,s}\,$.}
  \label{fig:Iter}
\end{figure}

\begin{figure}
  \centering
  \subfigure[$\eta-$profile]{\includegraphics[width=0.49\textwidth]{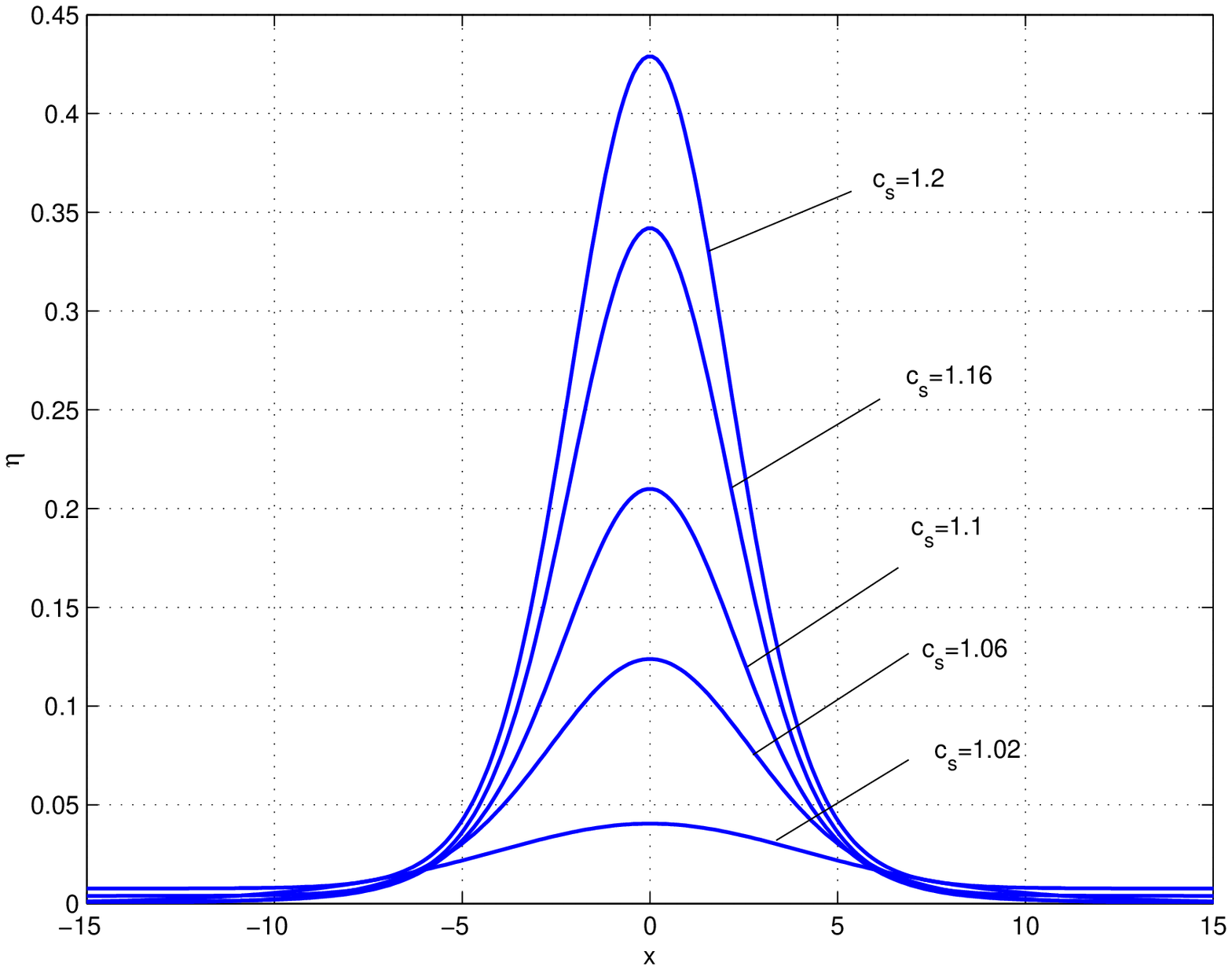}}
  \subfigure[$u-$profile]{\includegraphics[width=0.49\textwidth]{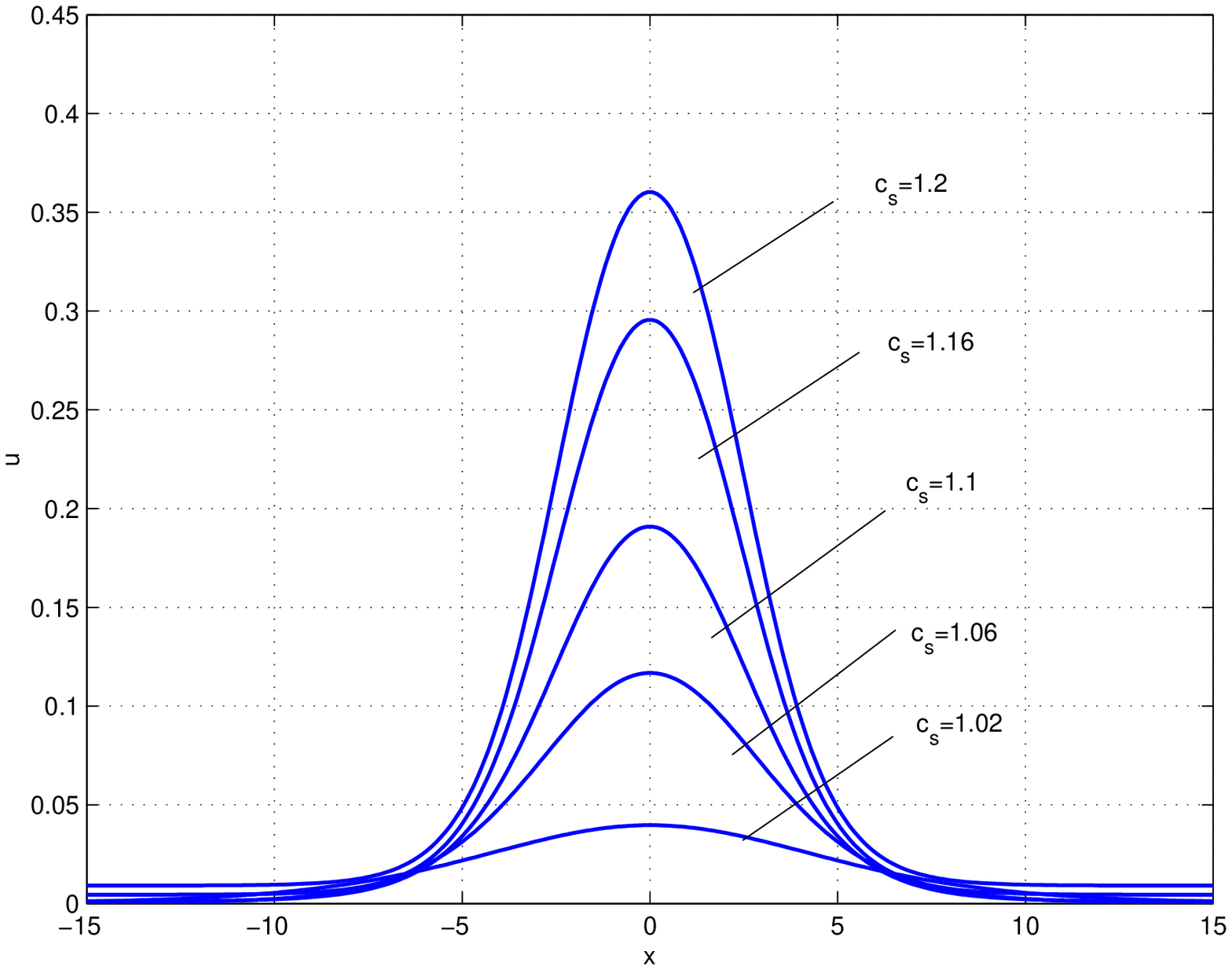}}
  \caption{\small\em Solitary wave profiles for various values of the propagation speed $c_{\,s}$ are superposed on the same image to show the evolution of the shape while changing this parameter. On the left image we show the free surface profile, while the right image depicts the horizontal velocity variable. The lowest curve corresponds to the smallest values of $c_{\,s}\ =\ 1.02$ and the highest solution is obtained for $c_{\,s}\ =\ 1.2\,$.}
  \label{fig:SWevol}
\end{figure}

The behaviour of the relative error $\eps_{\,1}\,[\nu]$ and absolute error $\eps_{\,2}\,[\nu]$ during the iterations is shown in Figure~\ref{fig:err} for two values of the propagation velocity $c_{\,s}\ =\ 1.05$ and $1.1\,$. In both cases, the iterations are stopped since the first error drops below the prescribed tolerance. The errors in Figure~\ref{fig:err} are measured in the maximum ($l_{\,\infty}$) norm. The results in the \textsc{Euclidean} ($l_{\,2}$) norm are completely similar. We can see that a relatively small number of iterations is needed to achieve the convergence. However, higher values of the propagation speed $c_{\,s}$ lead to higher nonlinearities. Consequently, more iterations are needed until the convergence is attained. The dependence of the number of iterations on the speed value $c_{\,s}$ is illustrated in Figure~\ref{fig:Iter}. The metamorphosis of these profiles as we change gradually the propagation speed $c_{\,s}$ is shown in Figure~\ref{fig:SWevol}.

\begin{figure}
  \centering
  \subfigure[Free surface]%
  {\includegraphics[width=0.49\textwidth]{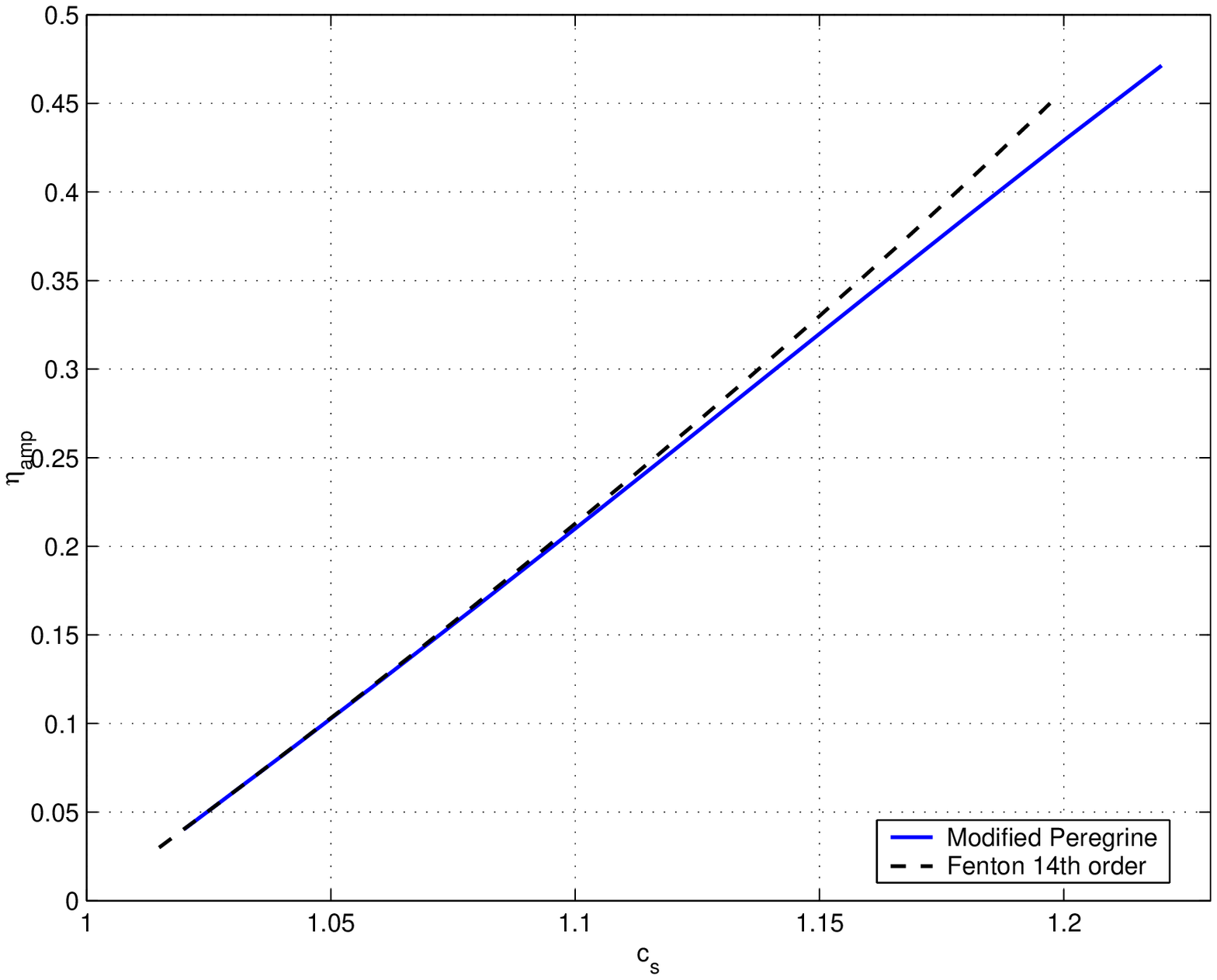}}
  \subfigure[Horizontal velocity]%
  {\includegraphics[width=0.49\textwidth]{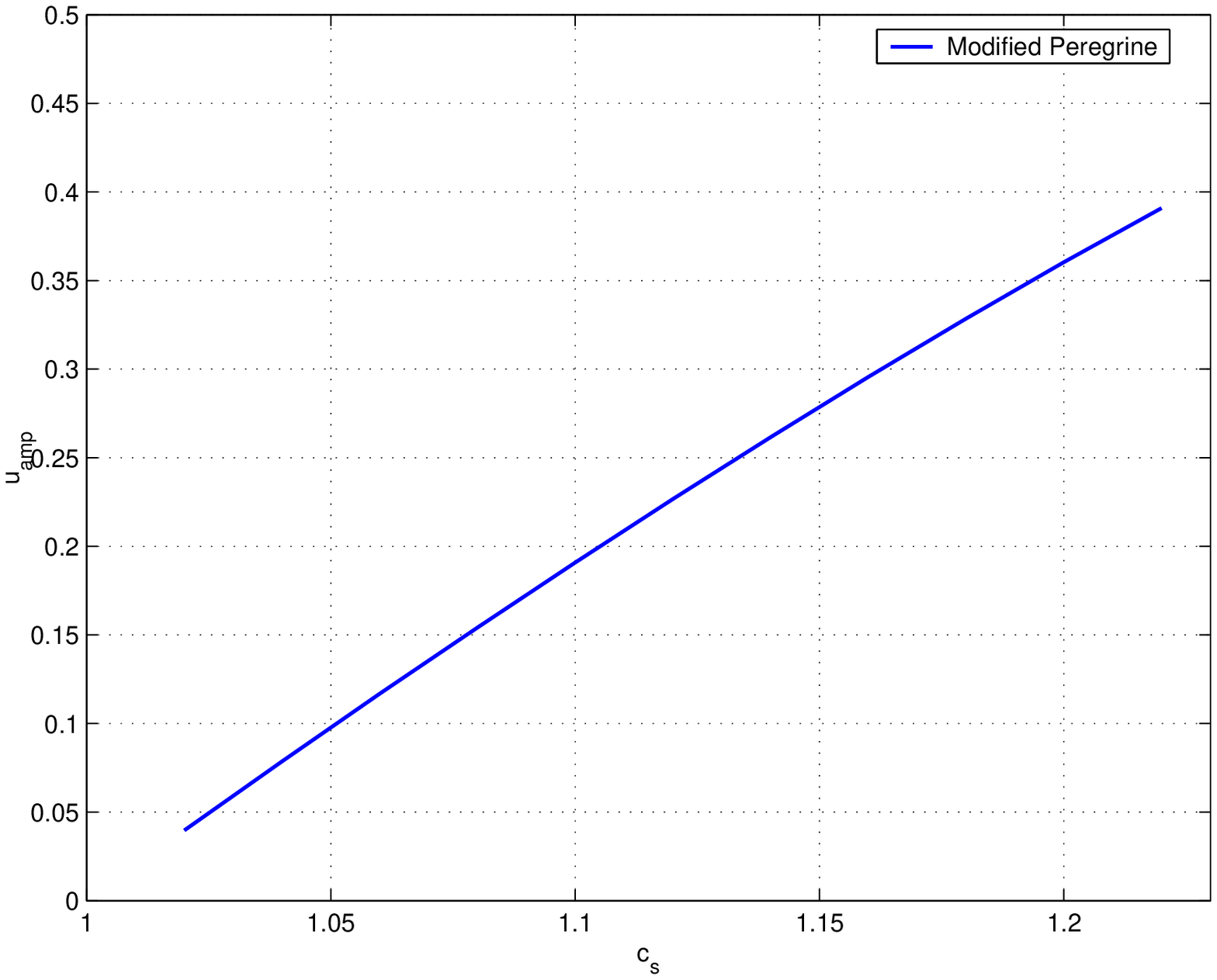}}
  \caption{\small\em Speed-amplitude relation for the m-\textsc{Peregrine} system. On the left image we show the free surface elevation amplitude and compare it to the 14\up{th} order \textsc{Fenton}'s solution. On the right image we show the horizontal velocity amplitude as a function of the propagation speed $c_{\,s}\,$.}
  \label{fig:ampcs}
\end{figure}

For illustrative purposes we provide several computed amplitudes (free surface elevation and horizontal velocity) of the solitary waves for various values of the propagation speed $c_{\,s}\,$. This speed-amplitude relation is represented graphically in Figure~\ref{fig:ampcs}. We make also a comparison with the $14$\up{th} order \textsc{Fenton}'s solution for the full water wave problem (for more details see \cite{Fenton1972, Longuet-Higgins1974}). One can notice a good agreement with the m-\textsc{Peregrine} system proposed in the previous Section.

\section{Numerical discretization}
\label{sec:fv}

In this section we present briefly the rationale on numerical methods we use to discretize the system \eqref{eq:cons1}, \eqref{eq:cons2} we derived above: Below we follow the great lines of our previous work \cite{Dutykh2011e}.


\subsection{Finite volume scheme}

We begin our presentation by a discretization of the hyperbolic part of equations (which are simply the classical nonlinear shallow water equations) and then, in the second time, we discuss the treatment of dispersive terms. The modified \textsc{Peregrine} system \eqref{eq:cons1}, \eqref{eq:cons2} can be formally put under this quasilinear form:
\begin{equation}\label{eq:conslaw}
  \D\,(v_{\,t})\ +\ [\,f\,(v)\,]_{\,x}\ =\ s\,(v)\,,
\end{equation}
where $v\,$, $f\,(v)$ are the conservative variables and the advective flux function respectively:
\begin{equation*}
  v\ =\ \begin{pmatrix}
        H \\
        Q
      \end{pmatrix}\,, \quad
  f\,(v)\ =\ \begin{pmatrix}
           Q \\
           \dfrac{Q^{\,2}}{H}\ +\ \dfrac{g}{2}\;H^{\,2}
         \end{pmatrix}\,.
\end{equation*}
The source term $s\,(v)$ contains the topography effects and $\D\,(v_{\,t})$ is the dispersion:
\begin{equation*}
  s\,(v)\ =\ \begin{pmatrix}
           0 \\
           g\, H\, h_{\,x}
         \end{pmatrix}\,, \quad
  \D\,(v_{\,t})\ =\ \begin{pmatrix}
             H_{\,t} \\
             \Bigl(1\ +\ \frac13\; H_{\,x}^{\,2}\ -\ \frac16\; H\,H_{\,x\,x}\Bigr)\,Q_{\,t}\ -\ \frac13\;H^{\,2}\,Q_{\,x\,x\,t}\ -\ \frac13\;H\,H_{\,x}\,Q_{\,x\,t}
           \end{pmatrix}\,.
\end{equation*}
Since the time derivative of the horizontal momentum $Q$ is defined implicitly, we will have to invert a linear elliptic operator with non-constant coefficients.

The \textsc{Jacobian} of the advective flux $f\,(v)$ can be easily computed:
\begin{equation*}
  \A\,(v)\ =\ \pd{f\,(v)}{v}\ =\ 
  \begin{pmatrix}
    0 & 1 \\
    g\,H - \Bigl(\displaystyle{\frac{Q}{H}}\Bigr)^{\,2} & \displaystyle{\frac{2\,Q}{H}}
  \end{pmatrix}\,.
\end{equation*}
The \textsc{Jacobian} $\A\,(v)$ has two distinctive eigenvalues:
\begin{equation*}
  \lambda^{\,\pm}\ =\ \frac{Q}{H}\ \pm\ c_{\,s}\ \equiv\ u\ \pm\ c_{\,s}\,, \qquad c_{\,s}\ \eqdef\ \sqrt{g\,H}\,.
\end{equation*}
The corresponding right and left eigenvectors are provided here:
\begin{equation*}
  R\ =\ \begin{pmatrix}
        1 & 1 \\
        \lambda^{\,+} & \lambda^{\,-}
      \end{pmatrix}\,, \qquad
  L\ =\ R^{\,-1}\ =\ -\frac{1}{2\,c_{\,s}}
            \begin{pmatrix}
                 \lambda^{\,-} & -1 \\
                 -\lambda^{\,+} & 1
               \end{pmatrix}\,.
\end{equation*}

Let us fix a partition of $\R$ into cells (or finite volumes) $\C_{\,i}\ =\ \bigl[\,x_{\,i\,-\,\frac12},\, x_{\,i\,+\,\frac12}\,\bigr]$ with cell centers $x_{\,i}\ =\ \frac{1}{2}\;(x_{\,i\,-\,\frac12}\ +\ x_{\,i\,+\,\frac12})\,$, $i\ \in\ \Z\,$. Let $\Delta x_{\,i}$ denotes the length of the cell $\C_{\,i}\,$. Without any loss of generality we assume the partition to be uniform, \ie $\Delta x_{\,i}\ \equiv\ \Delta x\,$, $\forall i\ \in\ \Z\,$. We would like to approximate the solution $v\,(x,\,t)$ by discrete values. In order to do so, we introduce the cell average of $v$ on the cell $\C_{\,i}$, \ie
\begin{equation*}
  \vm_{\,i}\,(t) \eqdef \bigl(\Hm_{\,i}\,(t),\, \Qm_{\,i}\,(t)\bigr)\ =\ \frac{1}{\Delta x}\; \int_{\C_{\,i}} v\,(x,\,t)\;\ud x\,.
\end{equation*}
A simple integration of \eqref{eq:conslaw} over the cell $\C_{\,i}$ leads the following exact relation:
\begin{equation*}
  \D\,(\vm_{\,t})_{\,i}\ +\ \frac{1}{\Delta x}\;\Bigl(f\,(v\,(x_{\,i\,+\,\frac12},\,t)\ -\ f\,(v\,(x_{\,i\,-\,\frac12},\,t))\Bigr)\ =\ \frac{1}{\Delta x}\;\int_{\C_{\,i}}\,s\,(v)\;\ud\,x\,.
\end{equation*}

Since the discrete solution is discontinuous at cell interfaces $x_{\,i\,+\,\frac12}\,$, $i\ \in\ \Z\,$, the heart of the matter in the finite volume method is to replace the flux through cell faces by the so-called numerical flux function:
\begin{equation*}
  f\,(v\,(x_{\,i\,\pm\,\frac12},\,t))\ \approx\ \F_{\,i\,\pm\,\frac12}\,(\vm_{\,i\,\pm\,\frac12}^{\,L}\,, \vm_{\,i\,\pm\,\frac12}^{\,R})\,,
\end{equation*}
where $\vm_{\,i\,\pm\,\frac12}^{\,L,\,R}$ are reconstructions of conservative variables $\vm$ from left and right sides of each cell interface. The reconstruction procedure employed in the present study will be described below. Consequently, the semi-discrete scheme takes the form:
\begin{equation}\label{eq:si1}
  \D\,(\vm_{\,t})_{\,i}\ +\ \frac{1}{\Delta x}\;\bigl(\F_{\,i\,+\,\frac12}\ -\ \F_{\,i\,-\,\frac12}\bigr)\ =\ \S_{\,i}\,,
\end{equation}
where $\S_{\,i}\ \approx\ \frac{1}{\Delta x}\;\int_{\C_{\,i}}\,s\,(v)\;\ud\,x$ is an approximation of the topographic term on the right-hand side of \eqref{eq:cons2}. In the present study we employ the standard hydrostatic reconstruction \cite{Audusse2004a} to obtain a well-balanced scheme. The expression for matrix $\D$ will be detailed below in Section~\ref{sec:dispers}.

In order to discretize the advective flux $f\,(v)$ we use the FVCF scheme \cite{Ghidaglia1996}:
\begin{equation*}
  \F\,(v,\,w)\ =\ \frac{f(v)\ +\ f(w)}{2}\ -\ U\,(v,\,w)\;\frac{f\,(w)\ -\ f\,(v)}{2}\,.
\end{equation*}
The first part of the numerical flux is centered, the second part is the upwinding introduced through the \textsc{Jacobian} sign matrix $U\,(v,\,w)$ defined as:
\begin{equation*}
  U\,(v,\,w)\ =\ \sign\bigl(\A\,(\mu)\bigr)\,, \qquad
  \sign\,(\A)\ =\ R\cdot\diag(s^{\,+},\, s^{\,-})\cdot L\,, \qquad
  s^{\,\pm}\ \eqdef\ \sign(\lambda^{\,\pm})\,.
\end{equation*}
The average state $\mu\ =\ (\mu_{\,1}\,(v,\,w),\, \mu_{\,2}(v,\,w)$ between the left $v\ =\ (H_{\,i\,+\,\frac12}^{\,L},\, u_{\,i\,+\,\frac12}^{\,L})$ and the right $w\ =\ (H_{\,i\,+\,\frac12}^{\,R},\, u_{\,i\,+\,\frac12}^{\,R})$ states\footnote{We do not take here the conservative variables $(H,\, Q)$ since the reconstruction procedure is more accurate and robust in physical variables $(H,\,u)\,$.} is defined as the \textsc{Roe} average:
\begin{equation*}
  \mu_{\,1}\ =\ \frac{H_{\,i\,+\,\frac12}^{\,L}\ +\ H_{\,i\,+\,\frac12}^{\,R}}{2}\,, \qquad
  \mu_{\,2}\ =\ \frac{\sqrt{H_{\,i\,+\,\frac12}^{\,L}}\,u_{\,i\,+\,\frac12}^{\,L}\ +\ \sqrt{H_{\,i\,+\,\frac12}^{\,R}}\,u_{\,i\,+\,\frac12}^{\,R}}{\sqrt{H_{\,i\,+\frac12\,}^{\,L}}\ +\ \sqrt{H_{\,i\,+\,\frac12}^{\,R}}}\,.
\end{equation*}
After some simple algebraic computations one can find the following expression for the sign matrix $U\,(v,\,w)\,$:
\begin{equation*}
  U\,(v,\,w)\ =\ \frac{1}{2\,c}\;\begin{pmatrix}
   s^{\,-}(\mu_{\,2}\ +\ c)\ -\ s^{\,+}\,(\mu_{\,2}\ -\ c) & s^{\,+}\ -\ s^{\,-} \\
   (s^{\,+}\ -\ s^{\,-})\,(c^{\,2}\ -\ \mu_{\,2}^{\,2}) & s^{\,+}\,(\mu_{\,2}\ +\ c)\ -\ s^{\,-}\,(\mu_{\,2}\ -\ c)
  \end{pmatrix}\,,
\end{equation*}
with $c \eqdef \sqrt{g\,\mu_{\,1}}\,$. We reiterate again that the sign matrix $U$ is evaluated at the average state $\mu$ of left and right values.


\subsection{High order reconstruction}

In order to obtain a higher order scheme in space, we need to replace the piecewise constant data by a piecewise polynomial representation. This goal is achieved by various so-called reconstruction procedures such as MUSCL TVD \cite{Kolgan1975, Leer1979, Leer2006}, UNO \cite{HaOs}, ENO \cite{Harten1989}, WENO \cite{Xing2005} and many others. In our previous study on Boussinesq-type equations \cite{Dutykh2011e}, the UNO2 scheme showed a good performance with low dissipation in realistic propagation and run-up simulations.

\begin{remark}
In TVD schemes the numerical operator is required (by definition) not to increase the total variation of the numerical solution at each time-step. It follows that the value of an isolated maximum may only decrease in time which is not a good property for the simulation of coherent structures such as solitary waves. The non-oscillatory UNO2 scheme, employed in our study, is only required to diminish the \emph{number} of local extrema in the numerical solution. Unlike TVD schemes, UNO schemes are not constrained to damp the values of each local extremum at every time-step.
\end{remark}

The main idea of the UNO2 scheme is to construct a non-oscillatory piecewise-parabolic interpolant $Q\,(x)$ to a piecewise smooth function $v\,(x)$ (see \cite{HaOs} for more details). On each segment containing the face $x_{\,i\,+\,\frac12}\ \in\ [x_{\,i},\, x_{\,i\,+\,1}\,]\,$, the function $Q\,(x)\ =\ q_{\,i\,+\,\frac12}\,(x)$ is locally a quadratic polynomial and wherever $v\,(x)$ is smooth we have:
\begin{equation*}
  Q\,(x)\ -\ v\,(x)\ =\ \O(\Delta x^{\,3}), \qquad
  \od{Q}{x}\,(x\,\pm\, 0)\ -\ \od{v}{x}\ =\ \O(\Delta x^{\,2})\,.
\end{equation*}
Also $Q\,(x)$ should be non-oscillatory in the sense that the number of its local extrema does not exceed that of $v\,(x)\,$. Since $q_{\,i\,+\,\frac12}\,(x_{\,i})\ =\ \vm_{\,i}$ and $q_{\,i\,+\,\frac12}\,(x_{\,i\,+\,1})\ =\ \vm_{\,i\,+\,1}\,$, it can be written in the form:
\begin{equation*}
  q_{\,i\,+\,\frac12}\,(x)\ =\ \vm_{\,i}\ +\ d_{\,i\,+\,\frac12}\,v\cdot\frac{x\ -\ x_{\,i}}{\Delta x}\ +\ \half\; D_{\,i\,+\,\frac12}\,v\cdot\frac{(x\ -\ x_{\,i})(x\ -\ x_{\,i\,+\,1})}{\Delta x^{\,2}}\,,
\end{equation*}
where $d_{\,i\,+\,\frac12}\,v \eqdef \vm_{\,i\,+\,1}\ -\ \vm_{\,i}$ and $D_{\,i\,+\,\frac12}\,v$ is closely related to the second derivative of the interpolant since $D_{\,i\,+\,\frac12}\,v\ =\ \Delta x^{\,2}\,q^{\,\prime\prime}_{\,i\,+\,\frac12}\,(x)\,$. The polynomial $q_{\,i\,+\,\frac12}\,(x)$ is chosen to be one the least oscillatory between two candidates interpolating $v\,(x)$ at $(x_{\,i\,-\,1},\, x_{\,i},\, x_{\,i\,+\,1})$ and $(x_{\,i},\, x_{\,i\,+\,1},\, x_{\,i\,+\,2})\,$. This requirement leads to the following choice of $D_{\,i\,+\,\frac12}\,v\,$:
\begin{equation*}
  D_{\,i\,+\,\frac12}\,v \eqdef \minmod\bigl(D_{\,i}\,v,\, D_{\,i\,+\,1}\,v\bigr)\,,
\end{equation*}
with
\begin{equation*}
  D_{\,i}\,v\ =\ \vm_{\,i\,+\,1}\ -\ 2\,\vm_{\,i}\ +\ \vm_{\,i\,-\,1}\,, \quad
  D_{\,i\,+\,1}\,v\ =\ \vm_{\,i\,+\,2}\ -\ 2\,\vm_{\,i\,+\,1}\ +\ \vm_{\,i}\,,
\end{equation*}
and $\minmod\,(x,\,y)$ is the usual min mod function defined as:
\begin{equation*}
  \minmod\,(x,\,y)\ =\ \frac{1}{2}\;(\sign(x)\ +\ \sign(y))\cdot\min(\abs{x},\,\abs{y})\,.
\end{equation*}
To achieve the second order $\O(\Delta x^{\,2})$ accuracy it is sufficient to consider piecewise linear reconstructions in each cell. Let $L\,(x)$ denote this approximately reconstructed function which can be written in this form:
\begin{equation*}
  L\,(x)\ =\ \vm_{\,i}\ +\ s_{\,i}\cdot\frac{x\ -\ x_{\,i}}{\Delta x}\,, \qquad
  x\ \in\ \bigl[\,x_{\,i\,-\,\frac12},\, x_{\,i\,+\,\frac12}\,\bigr]\,.
\end{equation*}
To make $L\,(x)$ a non-oscillatory approximation we use the parabolic interpolation $Q\,(x)$ constructed below to estimate the slopes $s_{\,i}$ within each cell:
\begin{equation*}
  s_{\,i}\ =\ \Delta x\cdot\minmod\Bigl(\od{Q}{x}\,(x_{\,i}\ -\ 0),\, \od{Q}{x}\,(x_{\,i}\ +\ 0)\Bigr)\,.
\end{equation*}
In other words, the solution is reconstructed on the cells while the solution gradient is estimated on the dual mesh as it is often performed in more modern schemes \cite{Barth1994, Barth2004}. A brief summary of the UNO2 reconstruction can be also found in \cite{Dutykh2011e}.


\subsection{Dispersive terms treatment}
\label{sec:dispers}

In this section we explain how we treat the dispersive terms of the m-\textsc{Peregrine} system \eqref{eq:cons1}, \eqref{eq:cons2}. Here again, we follow in great lines our previous study \cite{Dutykh2011e}. The following second order $\O(\Delta x^{\,2})$ approximations are used to discretize the dispersive terms arising in matrix $\D\,(v_{\,t})\,$:
\begin{multline*}
  \frac{1}{\Delta x}\;\int\limits_{\C_{\,i}}\Bigl[\,1\ +\ \frac13\;H_{\,x}^{\,2}\ -\ \frac16\; H\,H_{\,x\,x}\,\Bigr]\,Q_{\,t}\,\ud\,x\ \approx\\ 
  \Bigl(1\ +\ \frac13\;\bigl(\frac{H_{\,i\,+\,1}\ -\ H_{\,i\,-\,1}}{2\,\Delta x}\bigr)^{\,2}\ -\ \frac16\;H_{\,i}\,\frac{H_{\,i\,+\,1}\ -\ 2\,H_{\,i}\ +\ H_{\,i\,-\,1}}{\Delta x^{\,2}}\Bigr)\,(Q_{\,t})_{\,i}\,,
\end{multline*}
\begin{equation*}
  \frac{1}{\Delta x}\;\int\limits_{\C_{\,i}}\frac13\;H\,H_{\,x}\,Q_{\,x\,t}\,\ud\,x\ \approx\ \frac13\; H_{\,i}\;\frac{H_{\,i\,+\,1}\ -\ H_{\,i\,-\,1}}{2\,\Delta x}\;\frac{(Q_{\,t})_{\,i\,+\,1}\ -\ (Q_{\,t})_{\,i\,-\,1}}{2\,\Delta x}\,,
\end{equation*}
\begin{equation*}
  \frac{1}{\Delta x}\;\int\limits_{\C_{\,i}}\frac13\;H^{\,2}\,Q_{\,x\,x\,t}\,\ud\,x\ \approx\ \frac13\; H_{\,i}^{\,2}\;\frac{(Q_{\,t})_{\,i\,+\,1}\ -\ 2\,(Q_{\,t})_{\,i}\ +\ (Q_{\,t})_{\,i\,-\,1}}{\Delta x^{\,2}}\,.
\end{equation*}
Given the previous discretizations we obtain the following semi-discrete scheme:
\begin{eqnarray}\label{eq:19a}
  \od{\Hm_{\,i}}{t}\ +\ \frac{1}{\Delta x}\;\bigl(\F_{\,i\,+\,\frac12}^{\,(1)}\
  -\ \F_{\,i\,-\,\frac12}^{\,(1)}\bigr)\ &=&\ 0\,, \\
  \L\,\od{\Qm_{\,i}}{t}\ +\ \frac{1}{\Delta x}\;\bigl(\F_{\,i\,+\,\frac12}^{\,(2)}\
  -\ \F_{\,i\,-\,\frac12}^{\,(2)}\bigr)\ &=&\ \St\,(\vm)\,. \label{eq:19b}
\end{eqnarray}
The matrix $\D$ defined above in Equation~\eqref{eq:si1} can be expressed in terms of the matrix $\L\,$:
\begin{equation*}
  \D\ \eqdef\ \begin{pmatrix}
          \I & 0 \\
           0 & \L
        \end{pmatrix}\,,
\end{equation*}
where $\I$ is the identity matrix.

Consequently, in order to obtain the fully discrete scheme from Equations \eqref{eq:19a}, \eqref{eq:19b} we have to invert a system of linear equations with the tridiagonal matrix $\L\,$. It can be done efficiently with linear complexity. We note that on dry cells the matrix $\L$ becomes simply the identity matrix since $H_{\,i}\ \equiv\ 0$ in that regions. We reiterate again that we do not switch off the dispersive terms at some empirically chosen depth. It is the wave propagation physics which governs the magnitude of dispersive terms and thus, will decide whether they are important or not.


\subsection{Time-stepping}

We assume that the linear system of equations is already inverted leading to a system of ODEs of the form:
\begin{equation*}
  \vm_{\,t}\ =\ \N\,(\vm,\, t)\,, \qquad \vm\,(0)\ =\ \vm_{\,0}\,.
\end{equation*}
In order to solve numerically the last system of equations, we apply the \textsc{Bogacki}--\textsc{Shampine} method proposed in \cite{Bogacki1989}. It is a \textsc{Runge}--\textsc{Kutta} scheme of the third order with four stages. It has an embedded second order method which is used to estimate the local error and thus, to adapt the time-step size. Moreover, the \textsc{Bogacki}--\textsc{Shampine} method enjoys the First Same As Last (FSAL) property so that it needs approximately three function evaluations per step. This method is also implemented in the \texttt{ode23} function in \texttt{Matlab} \cite{Shampine1997}. The one step of the \textsc{Bogacki}--\textsc{Shampine} method is given by:
\begin{eqnarray*}
  k_{\,1}\ &=&\ \N\,(\vm^{\,(n)},\, t_{\,n})\,, \\
  k_{\,2}\ &=&\ \N\,(\vm^{\,(n)}\ +\ \half\;\Delta t_{\,n}\, k_{\,1},\, t_{\,n}\ +\ \half\;\Delta t)\,, \\
  k_{\,3}\ &=&\ \N\,(\vm^{\,(n)})\ +\ \frth\;\Delta t_{\,n}\, k_{\,2},\, t_{\,n}\ +\ \frth\;\Delta t)\,, \\
  \vm^{\,(n+1)}\ &=&\ \vm^{\,(n)}\ +\ \Delta t_{\,n}\,\bigl(\textstyle{2\over9}\;k_{\,1}\ +\ \textstyle{1\over3}\;k_{\,2}\ +\ \textstyle{4\over9}\;k_{\,3}\bigr)\,, \\
  k_{\,4}\ &=&\ \N\,(\vm^{\,(n+1)},\, t_{\,n} + \Delta t_{\,n})\,, \\
  \vm_{\,2}^{\,(n+1)}\ &=&\ \vm^{\,(n)}\ +\ \Delta t_{\,n}\,\bigl(\textstyle{4\over{24}}\;k_{\,1}\ +\ \textstyle{1\over4}\;k_{\,2}\ +\ \textstyle{1\over3}\; k_{\,3} + \textstyle{1\over8}\;k_{\,4}\bigr)\,.
\end{eqnarray*}
Here $\vm^{\,(n)}\ \approx\ \vm\,(t_{\,n})\,$, $\Delta t$ is the time-step and $\vm_{\,2}^{\,(n+1)}$ is a second order approximation to the solution $\vm\,(t_{\,n\,+\,1})\,$, so the difference between $\vm^{\,(n+1)}$ and $\vm_{\,2}^{\,(n+1)}$ gives an estimation of the local error. The FSAL property consists in the fact that $k_{\,4}$ is equal to $k_{\,1}$ in the next time-step, thus saving one function evaluation.

If the new time-step $\Delta t_{\,n\,+\,1}$ is given by $\Delta t_{\,n\,+\,1}\ =\ \rho_{\,n}\,\Delta t_{\,n}\,$, then according to \texttt{H211b} digital filter approach \cite{Soderlind2003, Soderlind2006}, the proportionality factor $\rho_{\,n}$ is given by:
\begin{equation}\label{eq:tadapt}
  \rho_{\,n}\ =\ \Bigl(\frac{\delta}{\eps_{\,n}}\Bigr)^{\,\beta_{\,1}}\,\Bigl(\frac{\delta}{\eps_{\,n\,-\,1}}\Bigr)^{\,\beta_{\,2}}\;\rho_{\,n\,-\,1}^{\,-\alpha}\,,
\end{equation}
where $\eps_{\,n}$ is a local error estimation at time-step $t_{\,n}$ and constants $\beta_{\,1}\,$, $\beta_{\,2}$ and $\alpha$ are defined as:
\begin{equation*}
  \alpha\ =\ \frac14\,, \quad \beta_{\,1}\ =\ \frac{1}{4\,p}\,, \quad \beta_{\,2}\ =\ \frac{1}{4\,p}\,.
\end{equation*}
The parameter $p$ is the order of the scheme and $p\ =\ 3$ in our case.

\begin{remark}
The adaptive strategy \eqref{eq:tadapt} can be further improved if we regularize the factor $\rho_{\,n}$ before computing the next time-step $\Delta t_{\,n\,+\,1}\,$:
\begin{equation*}
  \Delta t_{\,n\,+\,1}\ =\ \hat\rho_{\,n}\,\Delta t_{\,n}\,, \qquad
  \hat\rho_{\,n}\ =\ \omega\,(\rho_{\,n})\,.
\end{equation*}
The function $\omega\,(\rho)$ is called \emph{the time-step limiter} and should be smooth, monotonically increasing and should satisfy the following conditions:
\begin{equation*}
  \omega\,(0)\ <\ 1\,, \quad \omega\,(+\infty)\ >\ 1\,, \quad \omega\,(1)\ =\ 1\,, \omega^{\,\prime}\,(1)\ =\ 1\,.
\end{equation*}
One possible choice was suggested in \cite{Soderlind2006}:
\begin{equation*}
  \omega\,(\rho)\ =\ 1\ +\ \kappa\,\arctan\Bigl(\frac{\rho\ -\ 1}{\kappa}\Bigr)\,.
\end{equation*}
In our computations the parameter $\kappa$ is set to $1\,$.
\end{remark}

Several validations of the above presented numerical scheme, including the convergence tests, run-up simulations as well as the comparison with experimental data \cite{Synolakis1987, Zelt1991} can be found in our previous numerical study \cite{Dutykh2011e}. Here we make a step forward in the application of the proposed numerical model to practical coastal engineering problems.


\section{Numerical results}
\label{sec:nums}

Using the numerical method described in the preceding section, we can perform some simulations of the wave run-up onto a plane beach. Consider a setup schematically depicted in Figure~\ref{fig:sketch}. The bathymetry defined on a segment $\bigl[\,a,\, c\,\bigr]$ is composed of two regions: constant depth region $z\ =\ -d_{\,0}\,$, for $x\ \in\ \bigl[\,a,\, b\,\bigr]$ and the constant slope region $z\ =\ -d_{\,0}\ +\ x\,\tan(\delta)\,$, $x\ \in\ \bigl[\,b,\, c\,\bigr]\,$. We will solve numerically a Boundary Value Problem (BVP). Namely, on the right end ($x\ =\ c$) we impose the wall boundary condition $\left.u\right\vert_{\,x\,=\,c}\ =\ 0\,$, while on the left boundary ($x\ =\ a$) we are given by the incident wave height. In the present study we will consider the run-up of a monochromatic periodic wave entering from the left side (see Figure~\ref{fig:sketch}):
\begin{equation*}
  H_{\,0}\,(t)\ =\ d_{\,0}\ +\ A\,\sin(\omega\,t)\,.
\end{equation*}
The computational domain is discretized into $N\ =\ 500$ equal control volumes. The time-step value is automatically chosen by the time-stepping algorithm. The values of various physical parameters are given in Table~\ref{tab:params}.

\begin{remark}
The rigorous imposing of an incident wave boundary condition in the context of various dispersive wave equations is essentially an open question. However, for the m-\textsc{Peregrine} system under consideration, we found an operational solution based on the hyperbolic part of these equations. The general method is described in \cite{Pascal2002}. The numerical flux through the first left face $x\ =\ a$ is found by considering incoming characteristics and is given by this formula:
\begin{equation*}
  \F\,(x\,=\,a,\, t)\ =\ \begin{pmatrix}
                 H_{\,0}\,(t)\,u_{\,0} \\
                 H_{\,0}\,(t)\,u_{\,0}^{\,2}\ +\ \frac{g}{2}\;H_{\,0}^{\,2}\,(t)
               \end{pmatrix}\,,
   \qquad
   u_{\,0}\ \eqdef\ u_{\,1}\ +\ \bigl(1\ -\ \frac{H_{\,1}}{H_{\,0}}\bigr)\,\sqrt{g\, H_{\,1}}\,,
\end{equation*}
where $(H_{\,1},\, u_{\,1})$ are the reconstructed physical variables on the left face from the fluid domain. Our numerical tests presented below demonstrate the robustness and efficiency of this approach.
\end{remark}

\begin{table}
  \centering
  \begin{tabular}{l|c}
  \hline\hline
  Undisturbed water depth, $d_0$ & 1 \\
  Gravity acceleration, $g$ & 1 \\
  Incident wave amplitude, $A$ & 0.3 \\
  Incident wave frequency, $\omega$ & 0.8 \\
  Final simulation time, $T$ & 29.0 \\
  Left boundary coordinate, $a$ & -8 \\
  Transition coordinate between regions, $b$~ & 0 \\
  Right boundary coordinate, $c$ & 16 \\
  Beach slope, $\tan(\delta)$ & 0.14 \\
  \hline\hline
  \end{tabular}
  \bigskip
  \caption{\small\em Values of various parameters used in convergence tests.}
  \label{tab:params}
\end{table}

\begin{figure}
  \centering
  \includegraphics[width=0.99\textwidth]{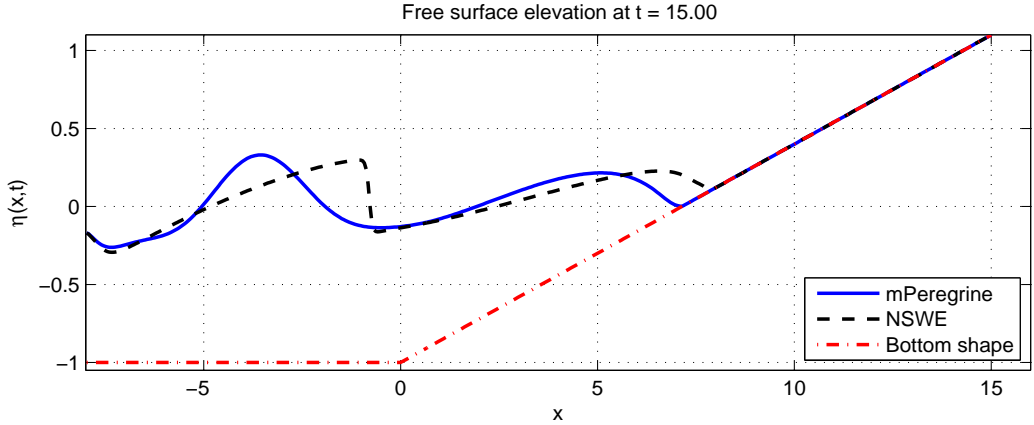}
  \caption{\small\em Free surface snapshot at $t\ =\ 15\,$. The blue solid line corresponds to the m-\textsc{Peregrine} system, the black dashed line refers to NSWE and the red dot-dashed line shows the bottom.}
  \label{fig:T15}
\end{figure}

\begin{figure}
  \centering
  \includegraphics[width=0.99\textwidth]{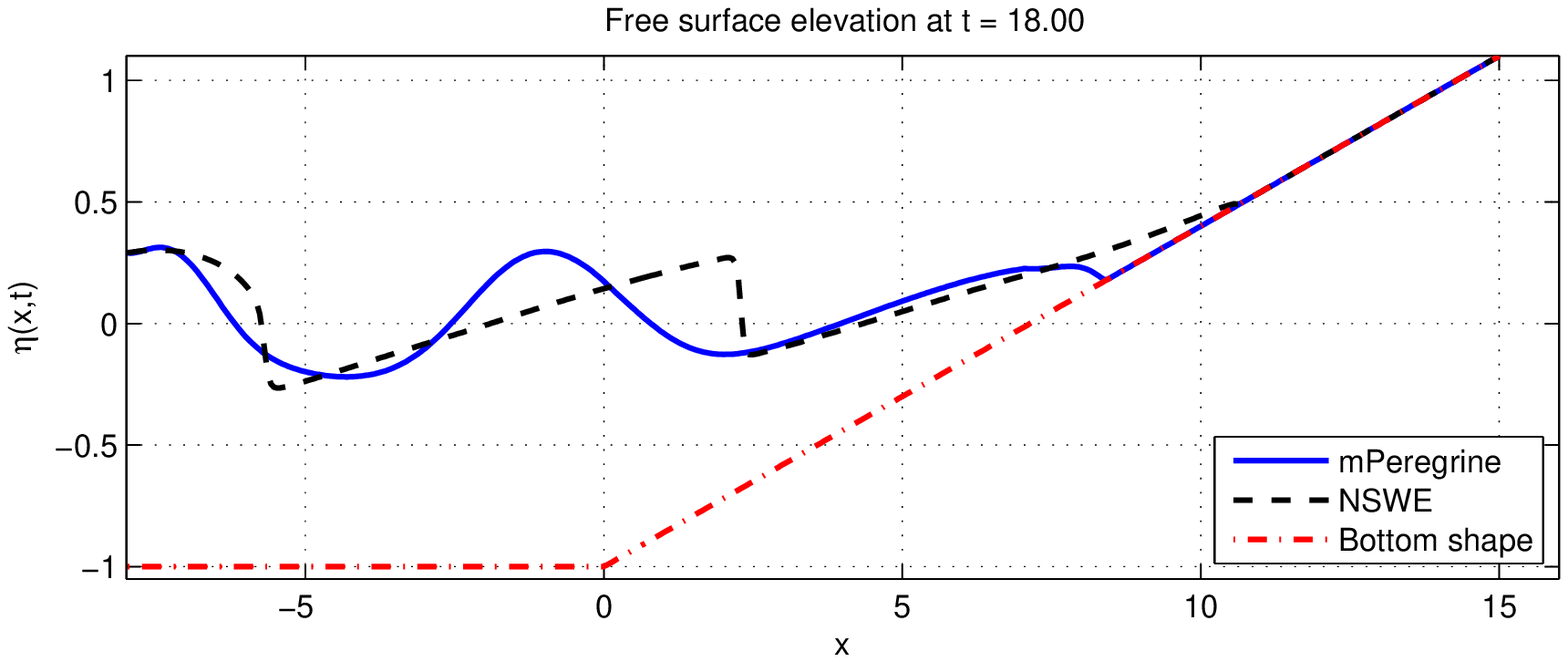}
  \caption{\small\em Free surface snapshot at $t\ =\ 18\,$. The blue solid line corresponds to the m-\textsc{Peregrine} system, the black dashed line refers to NSWE and the red dot-dashed line shows the bottom.}
  \label{fig:T18}
\end{figure}

\begin{figure}
  \centering
  \includegraphics[width=0.99\textwidth]{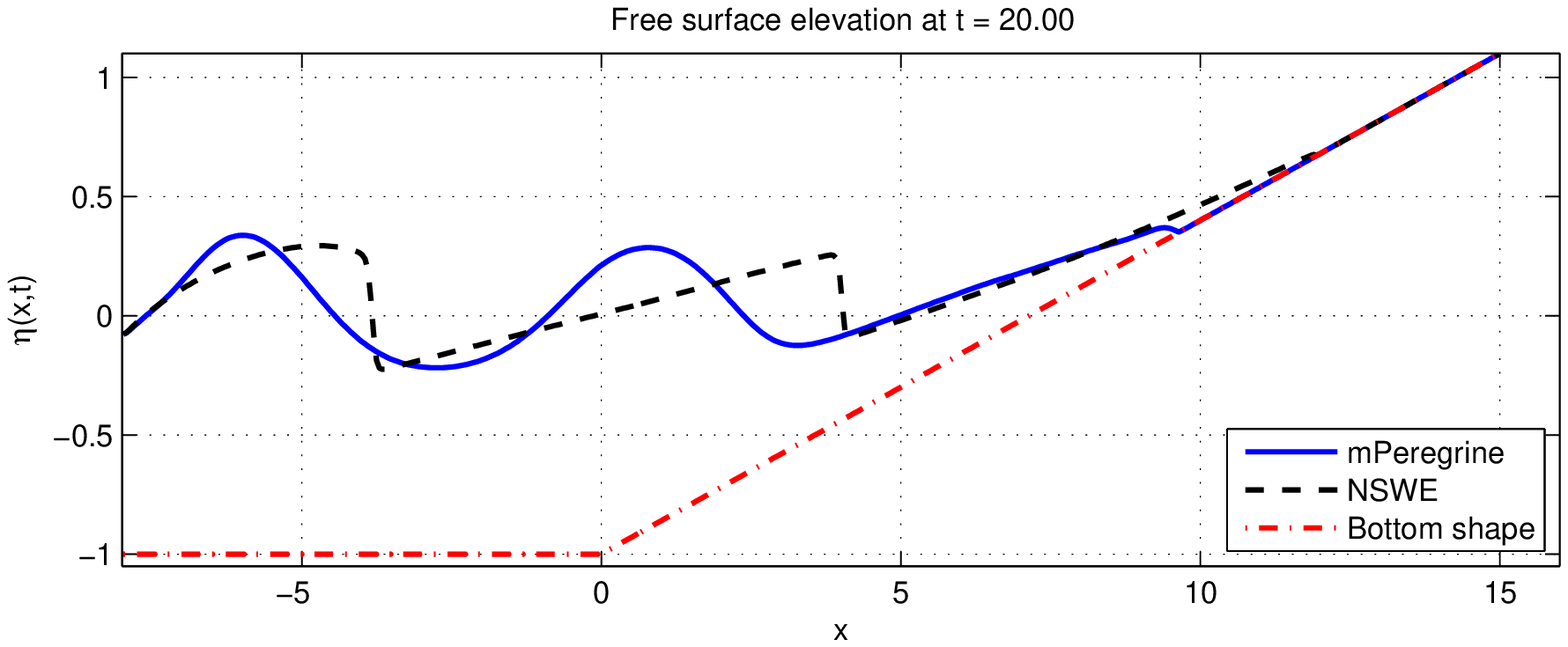}
  \caption{\small\em Free surface snapshot at $t\ =\ 20\,$. The blue solid line corresponds to the m-\textsc{Peregrine} system, the black dashed line refers to NSWE and the red dot-dashed line shows the bottom.}
  \label{fig:T20}
\end{figure}

\begin{figure}
  \centering
  \includegraphics[width=0.99\textwidth]{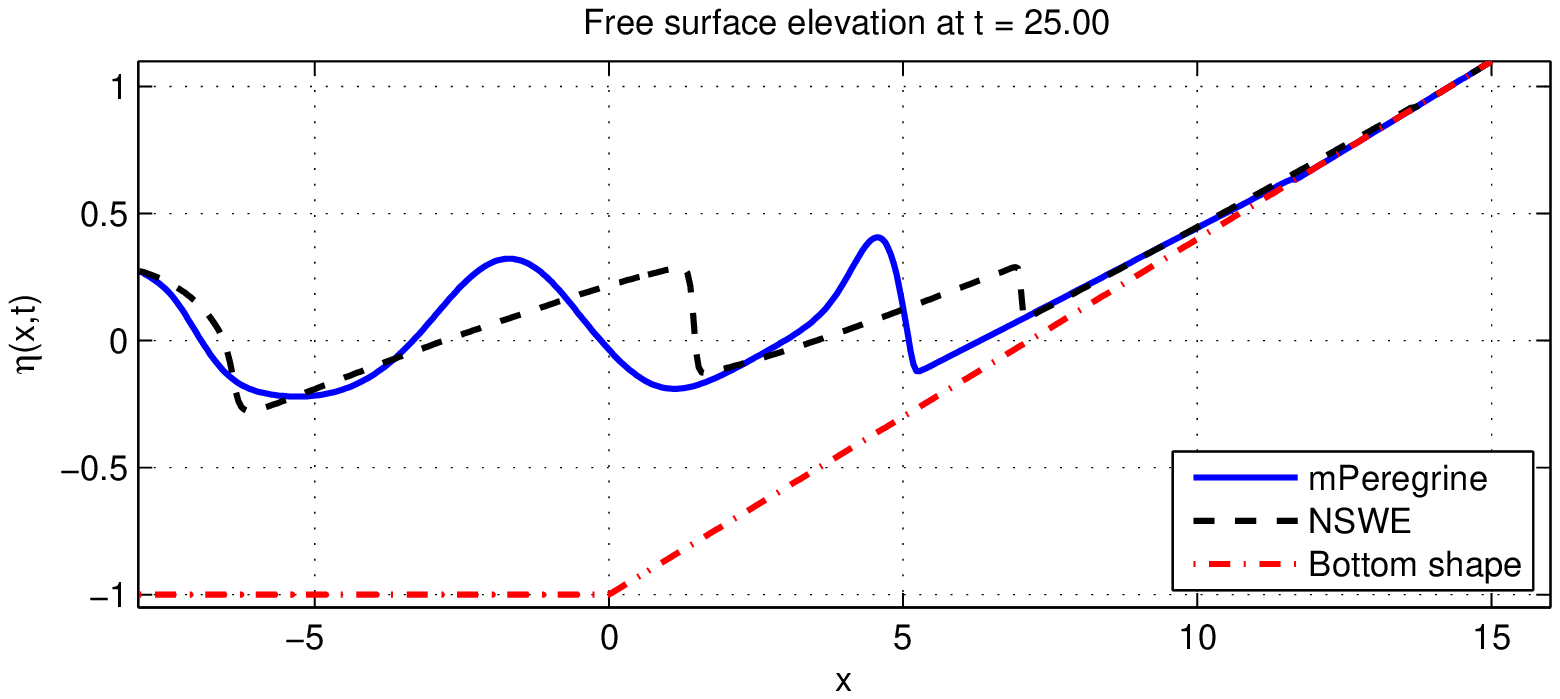}
  \caption{\small\em Free surface snapshot at $t\ =\ 25\,$. The blue solid line corresponds to the m-\textsc{Peregrine} system, the black dashed line refers to NSWE and the red dot-dashed line shows the bottom.}
  \label{fig:T25}
\end{figure}

The afore-described situation is simulated with the modified \textsc{Peregrine} system \eqref{eq:cons1}, \eqref{eq:cons2}, but also with classical nonlinear shallow water equations (NSWE) \cite{Zhou2002, Dutykh2009a, Dutykh2010c}. The comparative results of this simulation are presented in Figures \ref{fig:T15} -- \ref{fig:T25}. We underline that no friction terms are considered in this study. The numerical results we present are based only on mathematical models described above.

During the initial stages, which are not shown in figures for the sake of manuscript compactness, we see the periodic wave entering into the computational domain. The non-dispersive solution is much steeper and first shock waves start to form. Then, the wave continues its propagation towards the shore. During the propagation and run-up processes, the solution to the m-\textsc{Peregrine} system is always behind the hyperbolic wave and this is due to dispersive effects which make the wave propagation speed closer to its physical value. The run-up process starts about $t\ =\ 15$ and it can be seen in Figure~\ref{fig:T15}. The development of this process is shown in Figures \ref{fig:T18} -- \ref{fig:T25}. Both waves about their maximum run-up height are depicted in Figure~\ref{fig:T25}. It is interesting to observe a shock-like wave formed by the m-\textsc{Peregrine} system near the shore in Figure~\ref{fig:T25}. It shows that in the shallowest regions the wave dynamics is governed essentially by nonlinear effects. This transition is naturally and automatically captured by our system without adding any ad-hoc parameters.


\section{Landslide generated waves}
\label{sec:landslide}

Extreme water waves can become an important hazard in coastal areas. Main geophysical mechanisms include underwater earthquakes and landslides. The former genesis mechanism has been intensively investigated since the Tsunami Boxing Day \cite{Okal1988, Okal2003, Okal2004, Syno2006, Dutykh2006, Beisel2012, Dutykh2012, Dutykh2011d}. The list of references is far from being exhaustive. In this section we focus on the latter mechanism -- the underwater landslides which can cause some considerable damage in the genesis region. In general, the wavelength of landslide generated waves is much smaller than the length of transoceanic tsunamis. Consequently, the dispersive effects might be important. This consideration explains why we opt for a dispersive m-\textsc{Peregrine} model which is able to simulate the propagation and run-up of weakly nonlinear weakly dispersive water waves on nonuniform beaches.

Most of the landslide models which are currently used in the literature can be conventionally divided into three big categories. The first category contains the simplest models where the landslide shape and its trajectory are known {\em a priori} \cite{Tinti2001, todo2, MR2011541}. Another approach consists in assuming that the landslide motion is translational and the sliding mass follows the trajectory of its barycenter. The governing equation of the center of mass is obtained by projecting all the forces, acting on the slide, onto the horizontal direction of motion \cite{Grilli1999, Watts2000, DiRisio2009}. Finally, the third category of models describes the slide-water evolution as a two-layer system, the sliding mass being generally formulated by a \textsc{Savage}--\textsc{Hutter} type model \cite{Fernandez-Nieto2007}. Taking into account all the uncertainties which exist in the modeling of the real-world events, we choose in this chapter to study the intermediate level (\ie the second category) which corresponds better to the precision of the available data in real-world situations. The chosen landslide model will be detailed below in Section~\ref{sec:slide}.

The original derivation of the \textsc{Peregrine} system \cite{Peregrine1967} assumes that the bottom is stationary in time, \ie $z\ =\ -h\,(x)\,$. However, in order to simulate the wave generation process by bottom motion we need to include the time dependence into the bathymetry definition \cite{Dutykh2006, Dutykh2007b}. The bottom dynamics has been included into the \textsc{Peregrine} system derivation by T.~\textsc{Wu} \cite{Wu1981, Wu1987}: 
\begin{equation*}
  \eta_{\,t}\ +\ \bigl((h\ +\ \eta)\,u\bigr)_{\,x}\ =\ -\ h_{\,t}\,,
\end{equation*}
\begin{equation*}
  u_{\,t}\ +\ u\,u_{\,x}\ +\ g\,\eta_{\,x}\ -\ \frac{h}{2}\;(h\,u)_{\,x\,x\,t}\ +\ \frac{h^{\,2}}{6}\;u_{\,x\,x\,t}\ =\ \underbrace{\frac12\;h\,h_{\,x\,t\,t\,}}_{(*)}\,,
\end{equation*}
where the new term due to the bottom motion is marked with sign (*). By repeating the same invariantization process as above, the system written in conservative variables and with moving bottom can be straightforwardly derived:
\begin{equation}\label{eq:mper1}
  H_{\,t}\ +\ Q_{\,x}\ =\ 0\,,
\end{equation}
\begin{multline}\label{eq:mper2}
  \bigl(1\ +\ \frac13\; H_{\,x}^{\,2}\ -\ \frac16\;H\,H_{\,x\,x}\bigr)\,Q_{\,t}\ -\ \frac13\;H^{\,2}\,Q_{\,x\,x\,t}\ -\ \frac13\;H\,H_{\,x}\,Q_{\,x\,t}\ +\ \Bigl(\frac{Q^{\,2}}{H}\ +\ \frac{g}{2}\;H^{\,2}\Bigr)_{\,x}\\
  =\ g\,H\, h_{\,x}\ +\ \frac12\;H^2 \, h_{\,x\,t\,t}\,.
\end{multline}
The bottom motion enters into the momentum balance Equation~\eqref{eq:mper2} through the source term $\frac12\;H\,d_{\,x\,t\,t}\,$. The mass conservation Equation~\eqref{eq:mper1} keeps naturally its initial form. We underline that the linear dispersion relation of the m-\textsc{Peregrine} system \eqref{eq:mper1}, \eqref{eq:mper2} is identical with that the original \textsc{Peregrine} model \cite{Peregrine1967} since these models differ only in nonlinear terms and the source terms do not enter into the dispersion relation analysis. The numerical scheme described in Section~\ref{sec:fv} is applied to the moving bottom m-\textsc{Peregrine} system \eqref{eq:mper1}, \eqref{eq:mper2} without any modification. The new source term is just projected onto cell centers since the function $h\,(x,\,t)$ is prescribed by the bathymetry, the landslide shape and trajectory.

\begin{remark}
Following the same invariantization one can derive the two-di\-mensional modified \textsc{Peregrine} system including moving bottom topography:
\begin{align}
& H_{\, t}+\nabla \cdot  Q =0\\
&Q_{\, t}+\nabla \cdot \left(\, Q\, \otimes\,  \frac{Q}{H}\ +\ \frac{g}{2}\; H^{\, 2} \ {\bf {\rm I}}\, \right)\ -\ P\ (\, H,\ Q\, )\ =\ g\ H\ \nabla\  h\ +\ \frac{H^2}{2}\nabla h_{\, tt}\, ,
\end{align}
where 
$$P(\, H,\ Q\, ) \ = \ \frac{H^{\,2}}{2}\nabla(\nabla\cdot Q_{\, t})\ -\ \frac{H^2}{6}\Delta Q_{\, t}\ - \ \left(\, \frac{|\nabla H|^2}{3}-\frac{H\ \Delta H}{6}\, \right) \ Q_{\, t}\ + \ \frac{1}{3}\ H\ \nabla H\cdot \nabla Q_{\, t}\ .$$
It is noted that in this case $H$ depends on $(\, x,\ y,\ t\, )$ and $Q\ =\ H\, \times (u,v)^T$ with $u\, (\, x,\ y,\ t\, )$ and  $v\, (\, x,\ y,\ t\, )$ being the depth-averaged velocity horizontal components of the fluids velocity in the directions $x$ and $y$ respectively. This system again contains some high-order correction terms in the source terms that can be simplified without affecting the invariance of vertical translations.
\end{remark}


\subsection{Landslide model}
\label{sec:slide}

In this section we briefly present a model of an underwater landslide motion. This process has to be addressed carefully since it determines the subsequent formation of water waves. In this study we will assume the moving mass to be a solid quasi-deformable body with a prescribed shape and known physical properties that preserves its mass and volume. Under these assumptions it is sufficient to compute the trajectory of the barycenter $x\ =\ x_{\,c}\,(t)$ to determine the motion of the whole body. In general, only uniform slopes are considered in the literature in conjunction with this type of landslide models \cite{Pelinovsky1996, Grilli1999, Watts2000, DiRisio2009, Chubarov2011}. However, a novel model, taking into account the bottom geometry and curvature effects, has been recently proposed \cite{Beisel2012}. Hereafter we will follow in great lines this study.

The static bathymetry is prescribed by a sufficiently smooth (at least of the class $C^{\,2}$) and single-valued function $z\ =\ -h_{\,0}\,(x)\,$. The landslide shape is initially prescribed by a localized in space function $z\ =\ \zeta_{\,0}\,(x)\,$. For example, in this study we choose the following shape function:
\begin{equation}\label{eq:zeta}
  \zeta_{\,0}\,(x)\ =\ A\,\sech\bigl(k\,(x\ -\ x_{\,0})\bigr)\,,
\end{equation}
where the parameter $A$ is the maximum slide height, $k$ is inversely proportional to the slide length and $x_{\,0}$ is the initial position of its barycenter. Obviously, the model description given below is valid for any other reasonable shape.

Since the landslide motion is translational, its shape at time $t$ is given by the function $z\ =\ \zeta\,(x,\,t)\ =\ \zeta_{\,0}\,(x\ -\ x_{\,c}\,(t))\,$. Recall that the landslide center is located at the point with abscissa $x\ =\ x_{\,c}\,(t)\,$. Then, the impermeable bottom for the water wave problem can be easily determined at any time by simply superposing the static and dynamic components:
\begin{equation*}
  z\ =\ -h\,(x,\,t)\ =\ -h_{\,0}\,(x)\ +\ \zeta\,(x,\,t)\,.
\end{equation*}

To simplify the subsequent presentation, we introduce the classical arc-length parametrization, where the parameter $s\ =\ s\,(x)$ is given by the following formula:
\begin{equation}\label{eq:len}
  s\ =\ L\,(x)\ =\ \int_{x_{\,0}}^{x}\sqrt{1\ +\ (h_{\,0}^{\,\prime}(\xi))^{\,2}}\,\ud\,\xi\,.
\end{equation}
The function $L\,(x)$ is monotonic and can be efficiently inverted to turn back to the original \textsc{Cartesian} abscissa $x\ =\ L^{\,-1}\,(s)\,$. Within this parametrization, the landslide is initially located at point with the curvilinear coordinate $s\ =\ 0\,$. The local tangential direction is denoted by $\tau$ and the normal by $n\,$.

The landslide motion is governed by the following differential equation obtained by a straightforward application of \textsc{Newton}'s second law:
\begin{equation*}
  m\;\od{^2\,s}{t^{\,2}}\ =\ F_{\,\tau}\,(t)\,,
\end{equation*}
where $m$ is the mass and $F_{\,\tau}\,(t)$ is the tangential component of the forces acting on the moving submerged body. In order to project the forces onto the axes of local coordinate system, the angle $\theta\,(x)$ between $\tau$ and $O\,x$ can be easily determined:
\begin{equation*}
  \theta\,(x)\ =\ \arctan\bigl(h_{\,0}^{\,\prime}\,(x)\bigr)\,.
\end{equation*}

Let us denote by $\rho_{\,w}$ and $\rho_{\,\ell}$ the densities of the water and sliding material correspondingly. If $V$ is the volume of the slide, then the total mass $m$ is given by
\begin{equation*}
  m\ \eqdef\ (\rho_{\,\ell}\ +\ c_{\,w}\,\rho_{\,w})\,V\,,
\end{equation*}
where $c_w$ is the added mass coefficient \cite{Batchelor2000}. A portion of the water mass has to be added since it is entrained by the underwater body motion. The volume $V$ can be computed as
\begin{equation*}
  V\ =\ W\cdot S\ =\ W\,\int_{\R}\zeta_{\,0}\,(x) \, \ud\,x\,,
\end{equation*}
where $W$ is the landslide width in the transverse direction. The last integral can be computed exactly for the particular choice \eqref{eq:zeta} of the landslide shape to give
\begin{equation*}
  V\ =\ \frac12\;\ell\, A\, W\,.
\end{equation*}

The total projected force $F_{\,\tau}$ acting on the landslide can be conventionally represented as a sum of two different kind of forces denoted by $F_{\,g}$ and $F_{\,d}\,$:
\begin{equation*}
  F_{\,\tau}\ =\ F_{\,g}\ +\ F_{\,d}\,,
\end{equation*}
where $F_{\,g}$ is the joint action of the gravity and buoyancy, while $F_{\,d}$ is the total contribution of various dissipative forces (to be specified below). The gravity and buoyancy forces act in opposite directions and their horizontal projection $F_{\,g}$ can be easily computed:
\begin{equation*}
  F_{\,g}\,(t)\ =\ (\rho_{\,\ell}\ -\ \rho_{\,w})\,W\, g\,\int_\R \zeta\,(x,\,t)\,\sin\bigl(\theta\,(x)\bigr)\,\ud\,x\,.
\end{equation*}
Now, let us specify the dissipative forces. The water resistance to the motion force $F_{\,r}$ is proportional to the maximal transversal section of the moving body and to the square of its velocity:
\begin{equation*}
  F_{\,r}\ =\ -\frac12\; c_{\,d}\,\rho_{\,w}\, A\,W\,\sigma\,(t)\Bigl(\od{s}{t}\Bigr)^{\,2}\,,
\end{equation*}
here $c_{\,d}$ is the resistance coefficient of the water and $\sigma\,(t)\ \eqdef\ \sign\Bigl(\od{s}{t}\Bigr)\,$. The coefficient $\sigma\,(t)$ is needed to dissipate the landslide kinetic energy independently of its direction of motion. The friction force $F_{\,f}$ is proportional to the normal force exerted on the body due to the weight:
\begin{equation*}
  F_{\,f}\ =\ -c_{\,f}\,\sigma\,(t)\,N\,(x,\,t)\,.
\end{equation*}
The normal force $N\,(x,\,t)$ is composed of the normal components of gravity and buoyancy forces but also of the centrifugal force due to the variation of the bottom slope:
\begin{equation*}
  N\,(x,\,t)\ =\ (\rho_{\,\ell}\ -\ \rho_{\,w})\,g\,W\,\int_\R\zeta\,(x,\,t)\,\cos\bigl(\theta\,(x)\bigr)\,\ud\,x\ +\ \rho_{\,\ell}\,W\,\int_\R\zeta\,(x,\,t)\,\kappa\,(x)\Bigl(\od{s}{t}\Bigr)^{\,2}\, \ud\,x\,,
\end{equation*}
where $\kappa\,(x)$ is the signed curvature of the bottom which can be computed by the following formula:
\begin{equation*}
  \kappa\,(x)\ =\ \frac{h_{\,0}^{\,\prime\prime}\,(x)}{\bigl(1\ +\ (h_{\,0}^{\,\prime}\,(x))^{\,2}\bigr)^{\,\frac32}}\,.
\end{equation*}
We note that the last term vanishes for a plane bottom since $\kappa\,(x)\ \equiv\ 0$ in this particular case.

In order to dissipate more energy along the landslide trajectory if it is needed, we complete our model by two supplementary viscous terms:
\begin{equation*}
  F_{\,d}\ =\ -c_{\,v}\;\od{s}{t}\ -\ c_{\,b}\;\od{s}{t}\;\abs{\od{s}{t}}\,,
\end{equation*}
where $c_{\,v}$ and $c_{\,b}$ are some prescribed constants. The first term $c_{\,v}$ represents the internal energy loss inside the sliding material. The second term $c_{\,b}$ accounts for the dissipation in the boundary layer between the landslide and the solid bottom.

Finally, if we sum up all the contributions of described above forces, we obtain the following second order differential equation:
\begin{multline}\label{eq:ODE}
  (\gamma\ +\ c_{\,w})\,S\,\od{^2\,s}{t^{\,2}}\ =\ (\gamma\ -\ 1)\,g\,\Bigl(I_{\,1}\,(t)\ -\ c_{\,f}\,\sigma\,(t)\,I_{\,2}\,(t)\Bigr) \\
  -\ \sigma\,(t)\,\Bigl(c_{\,f}\,\gamma\,I_{\,3}\,(t)\ +\ \frac12\;c_{\,d}\,A\Bigr)\Bigl(\od{s}{t}\,\Bigr)^{\,2}\ -\ c_{\,v}\,\od{s}{t}\ -\ c_{\,b}\;\od{s}{t}\abs{\od{s}{t}}\,,
\end{multline}
where $\gamma\ \eqdef\ \frac{\rho_{\,\ell}}{\rho_{\,w}}\ >\ 1$ is the ratio of densities and integrals $I_{\,1,\,2,\,3}\,(t)$ are defined as:
\begin{align*}
  I_{\,1}\,(t)\ &=\ \int_\R\zeta\,(x,\,t)\,\sin\bigl(\theta\,(x)\bigr)\,\ud\,x\,,\\
  I_{\,2}\,(t)\ &=\ \int_\R\zeta\,(x,\,t)\,\cos\bigl(\theta\,(x)\bigr)\,\ud\,x\,, \\
  I_{\,3}\,(t)\ &=\ \int_\R\zeta\,(x,\,t)\,\kappa\,(x)\,\ud\,x\,.
\end{align*}
Note also that Equation~\eqref{eq:ODE} was simplified by dividing both sides by the width value $W\,$. In order to obtain a well-posed initial value problem, Equation~\eqref{eq:ODE} has to be completed by two initial conditions:
\begin{equation*}
  s\,(0)\ =\ 0\,, \quad s^{\,\prime}\,(0)\ =\ 0\,.
\end{equation*}
From Equation~\eqref{eq:ODE} it follows that the motion can start only if this condition is fulfilled \cite{Beisel2012}:
\begin{equation*}
  I_{\,1}\,(0)\ -\ c_{\,f}\, I_{\,2}\,(0)\ =\ \int_\R\,\zeta_{\,0}\,(x)\Bigl[\sin\bigl(\theta\,(x)\bigr)\ -\ c_{\,f}\cos\bigl(\theta\,(x)\bigr)\Bigr]\,\ud\,x\ >\ 0\,.
\end{equation*}

In order to solve numerically Equation~\eqref{eq:ODE} we employ the same \textsc{Bogacki}--\textsc{Shampine} 3\up{rd} order \textsc{Runge}--\textsc{Kutta} scheme that we used to approximate the \textsc{Boussinesq} Equations \eqref{eq:mper1}, \eqref{eq:mper2}. The integrals $I_{\,1,\,2,\,3}\,(t)$ are computed with the trapezoidal rule. Once the landslide trajectory $s\ =\ s\,(t)$ is found, Equation~\eqref{eq:len} is used to find its motion $x\ =\ x\,(t)$ in the initial \textsc{Cartesian} coordinate system.


\subsection{Numerical results}

\begin{figure}
  \centering
  \includegraphics[width=0.99\textwidth]{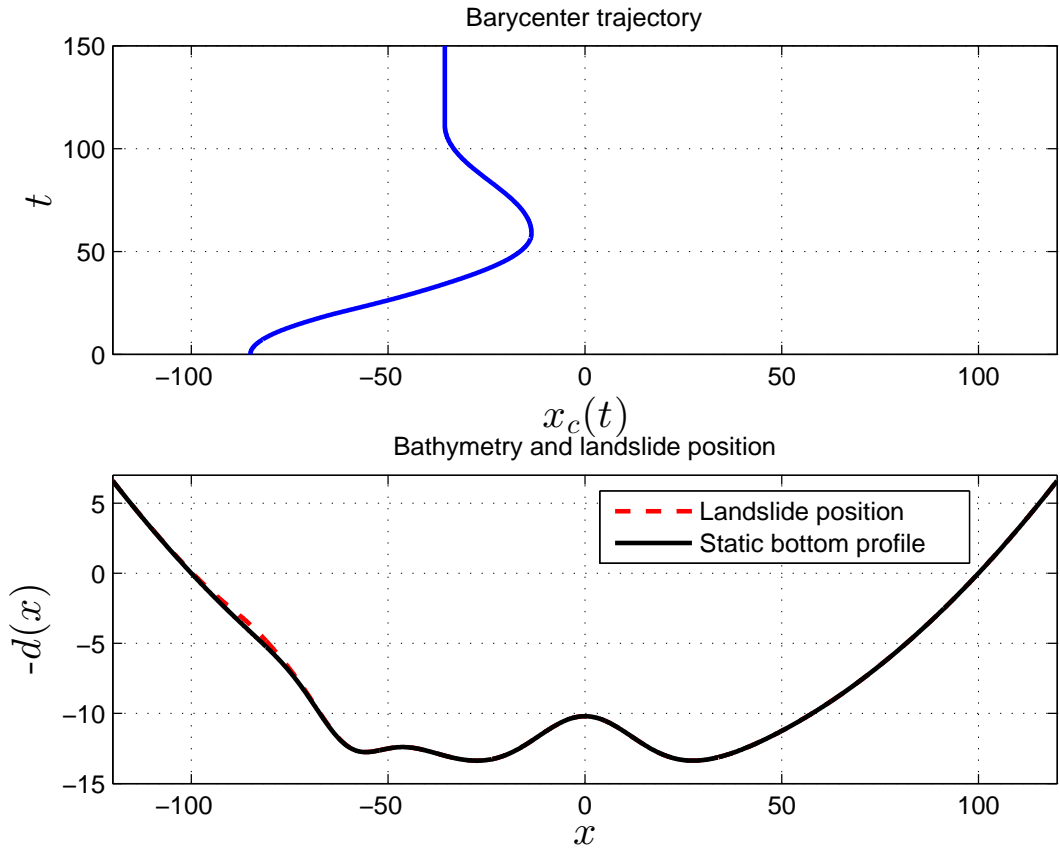}
  \caption{\small\em Bathymetry profile and the landslide trajectory for the parameters given in Table~\ref{tab:pars}. The initial landslide position is shown on the lower image with the red dashed line (\textcolor{red}{- - -}).}
  \label{fig:bot}
\end{figure}

\begin{figure}
  \centering
  \includegraphics[width=0.99\textwidth]{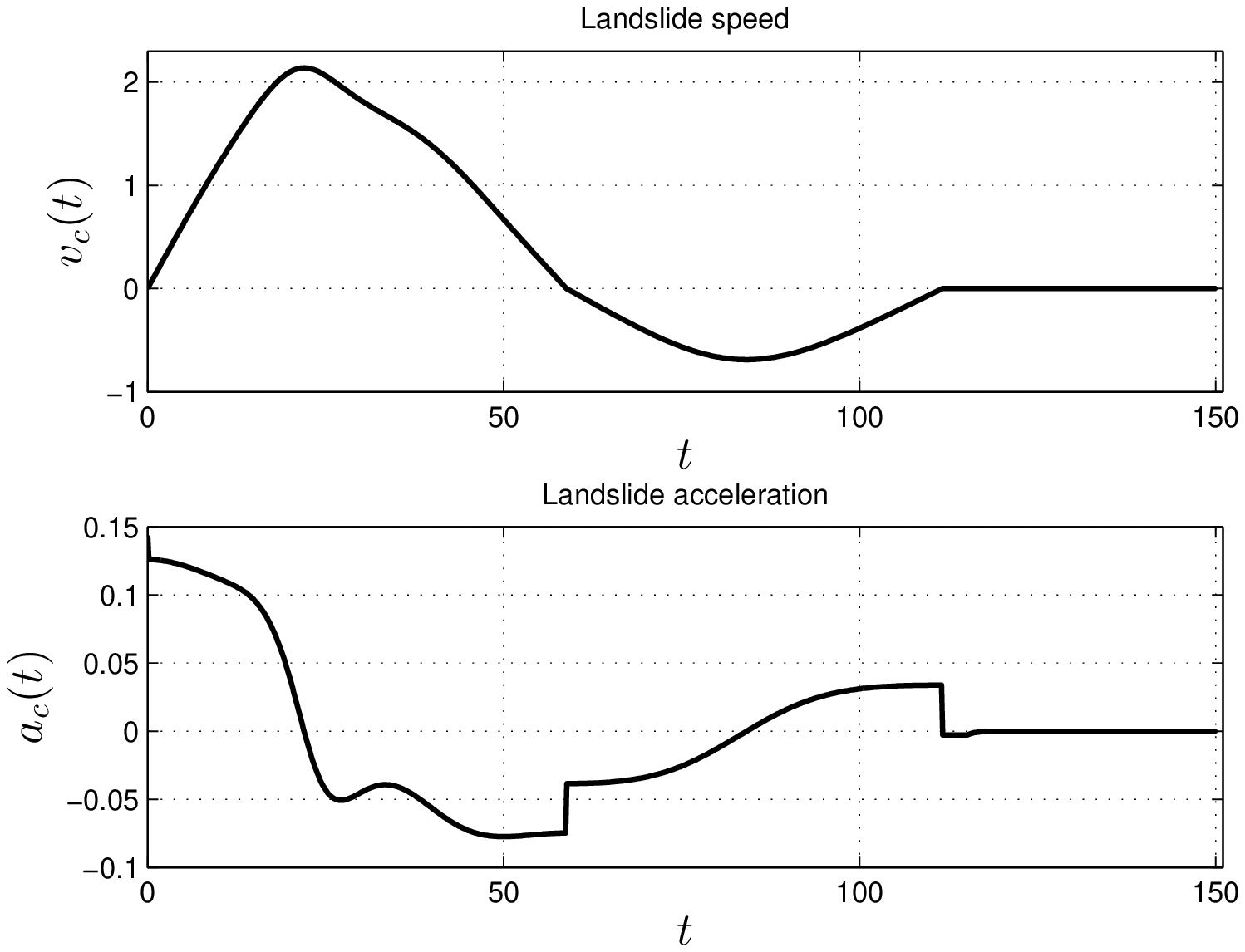}
  \caption{\small\em Landslide speed and acceleration along its trajectory.}
  \label{fig:SpAc}
\end{figure}

Consider a one-dimensional physical domain ${\rm I}\ =\ \bigl[\,a,\, b\,\bigr]\ =\ \bigl[\,-120,\, 120\,\bigr]$ which is divided into $N$ equal control volumes. This domain is composed of three regions: the left and right curvilinear sloping beaches which surround a generation region of a deformed parabolic shape. Specifically, the static bathymetry function $d_{\,0}\,(x)$ is given by the following expression:
\begin{equation*}
  d_{\,0}\,(x)\ =\ -\kappa\,\bigl(x^{\,2}\ -\ c^{\,2}\bigr)\ +\ A_{\,1}\,\ue^{-k_{\,1}\,(x\ -\ x_{\,1})^{\,2}}\ +\ A_{\,2}\,\ue^{-k_{\,2}\,(x\ -\ x_{\,2})^{\,2}}\,.
\end{equation*}
Basically, this function represents a parabolic bottom profile deformed by two underwater bumps. We made this nontrivial choice in order to illustrate better the advantages of our landslide model, which was designed to handle general non-flat bathymetries. The values of all physical and numerical parameters are given in Table~\ref{tab:pars}. The bottom profile along with landslide trajectory for these parameters are depicted in Figure~\ref{fig:bot}. The landslide motion starts from the rest position under the action of the gravity force. We simulate its motion along with the free surface waves up to time $T\ =\ 150.0$ \textsf{s}. As it is expected, the landslide remains trapped between two underwater bumps in its final equilibrium position. The speed and acceleration of the slide barycenter during the simulation are represented in Figure~\ref{fig:SpAc}. We note the discontinuities in the acceleration record which correspond to the time moments when the velocity changes its sign. We insist that this behaviour is intrinsic to the landslide model in use where the dissipative terms show the discontinuous behaviour at turning points.

\begin{table}
 \centering
 \begin{tabular}{c|c}
   \hline\hline
   \textit{Parameter} & \textit{Value} \\
   \hline\hline
   Gravity acceleration, $g$ & $1.0$ \\
   Parabolic bottom flatness coefficient, $\kappa$ & $1.5\times 10^{-3}$ \\
   Initial shoreline position, $c$ & $100.0$ \\
   Underwater bump amplitude, $A_{\,1}$ & $2.8$ \\
   Underwater bump amplitude, $A_{\,2}$ & $-4.8$ \\
   Bump characteristic steepness, $k_{\,1}$ & $0.008$ \\
   Bump characteristic steepness, $k_{\,2}$ & $0.003$ \\
   Bump center position, $x_{\,1}$ & $-60.0$ \\
   Bump center position, $x_{\,2}$ & $0.0$ \\
   Number of control volumes, $N$ & $2500$ \\
   Slide amplitude, $A$ & $0.5$ \\
   Characteristic slide inverse length, $k_{\,0}$ & $0.16$ \\
   Initial slide position, $x_{\,0}$ & $-85.0$ \\
   Added mass coefficient, $c_{\,w}$ & $1.0$ \\
   Water drag coefficient, $c_{\,d}$ & $1.0$ \\
   Friction coefficient, $c_{\,f}$ & $\tan2^\circ$ \\
   Ratio between water and slide densities, $\gamma$ & $2.0$ \\
   Boundary layer dissipation coefficient, $c_{\,b}$ & $0.0035$ \\
   Internal friction coefficient, $c_{\,v}$ & $0.0045$ \\
   Final simulation time, $T$ & $150.0$ \\
   \hline\hline
 \end{tabular}
 \bigskip
 \caption{\small\em Values of various parameters used in the numerical computations.}
 \label{tab:pars}
\end{table}

\begin{figure}
  \centering
  \includegraphics[width=0.9\textwidth]{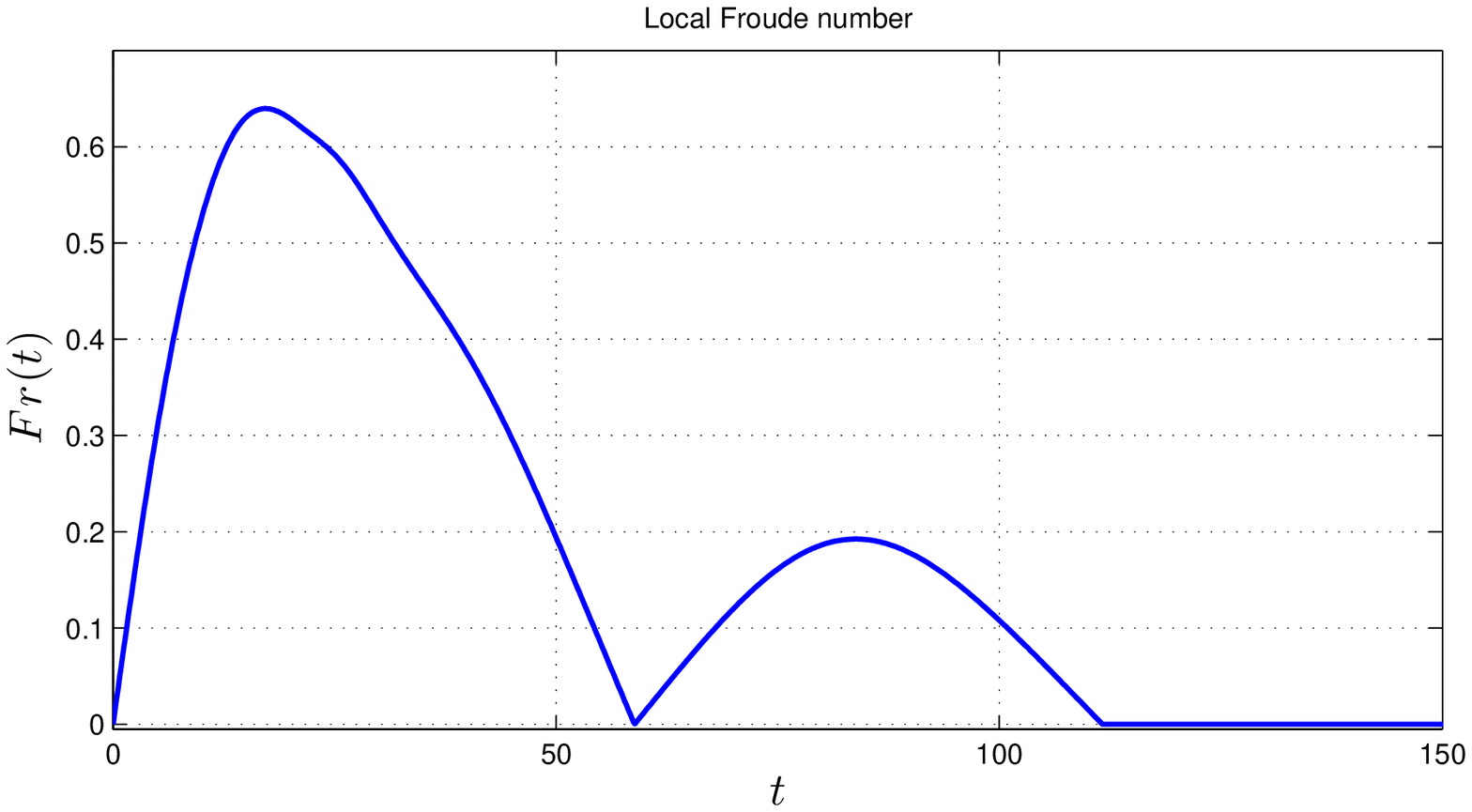}
  \caption{\small\em Local \textsc{Froude} number computed along the slide motion.}
  \label{fig:froude}
\end{figure}

One of the important parameters in shallow water flows is the \textsc{Froude} number, defined as the ratio between the characteristic fluid velocity to the gravity wave speed. We computed also this parameter along the landslide trajectory:
\begin{equation*}
  \mathrm{Fr}\,(t)\ \eqdef\ \frac{\abs{x_{\,c}^{\,\prime}\,(t)}}{\sqrt{g\,d\,\bigl(x_{\,c}\,(t),\,t)\bigr)}}\,.
\end{equation*}
The result is presented in Figure~\ref{fig:froude}. We can see that in our case the slide motion remains sub-critical as it is the case in most real world situations \cite{Harbitz2006}.

\begin{figure}
  \centering
  \includegraphics[width=0.99\textwidth]{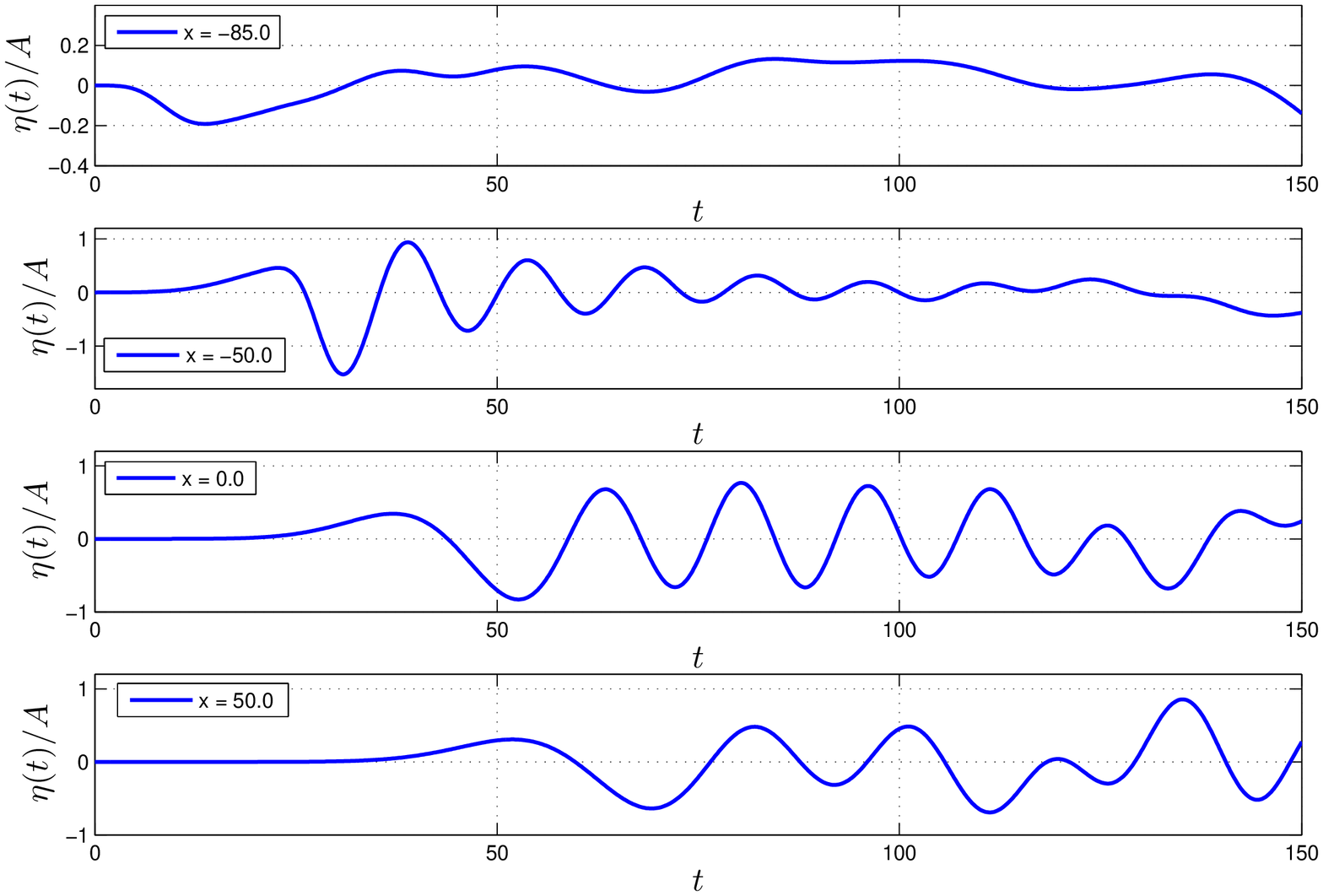}
  \caption{\small\em Synthetic wave gauge records at four different locations. Note the different vertical scales on various images. Wave gauges are located at $x\ =\ x_{\,0}\ =\ -85.0\,$, $-50.0\,$, $0.0\,$, $50.0$ from the top correspondingly. The wave amplitude is relative to the landslide amplitude.}
  \label{fig:gauges}
\end{figure}

\begin{figure}
  \centering
  \includegraphics[width=0.99\textwidth]{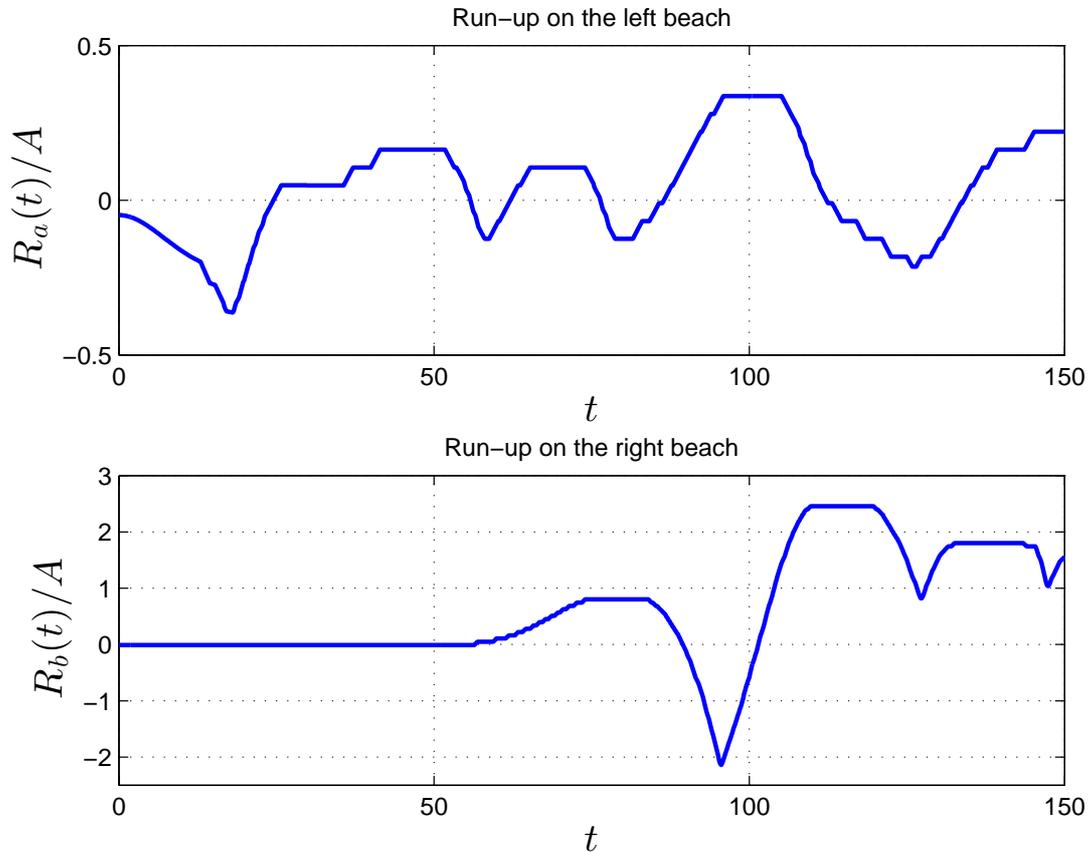}
  \caption{\small\em Wave run-up heights onto left and right non-flat beaches during the simulation.}
  \label{fig:runup}
\end{figure}

\begin{figure}
  \centering
  \includegraphics[width=0.99\textwidth]{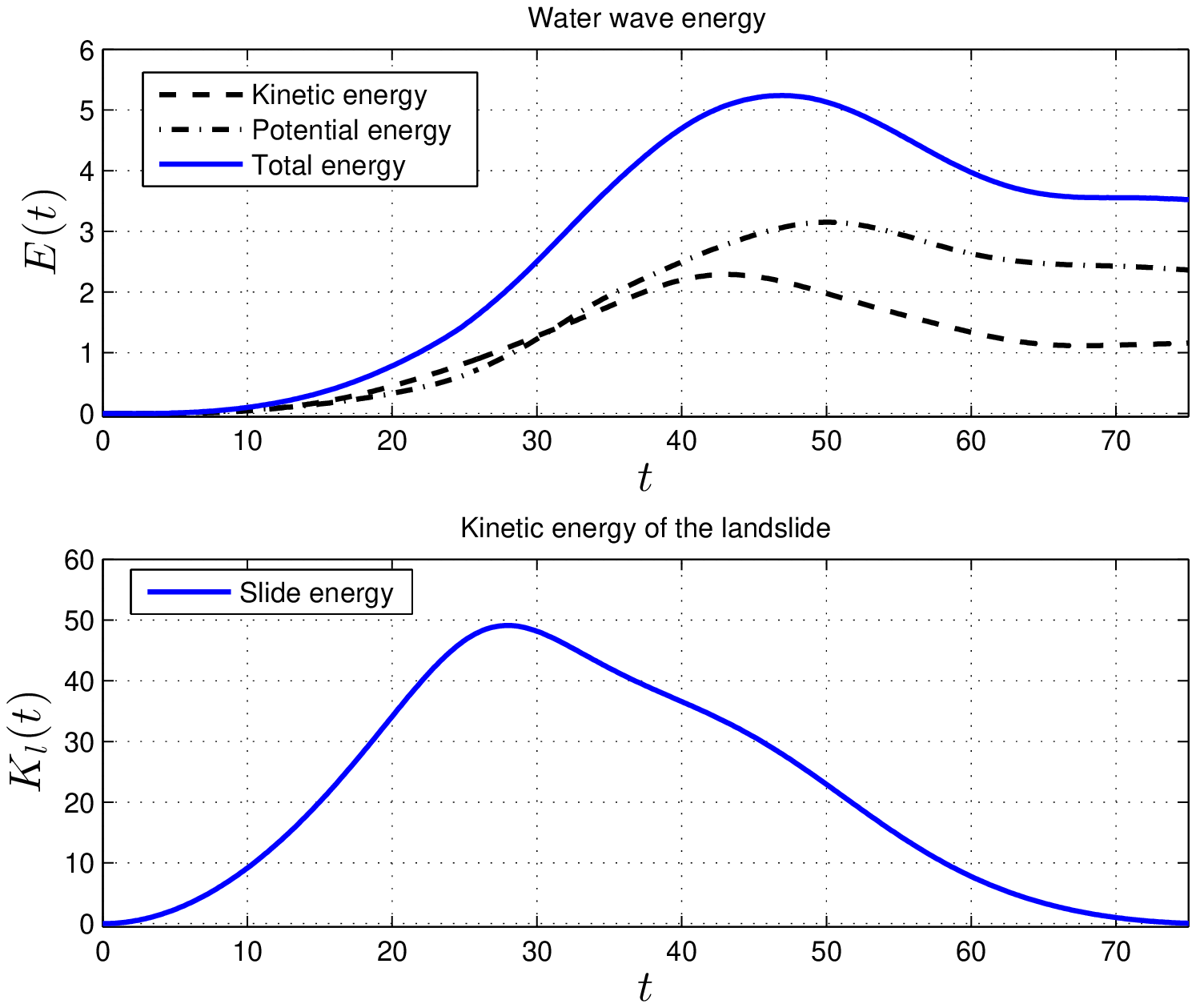}
  \caption{\small\em Fluid and landslide energies evolution during the wave generation process.}
  \label{fig:energy}
\end{figure}

In order to measure the free surface elevations due to the underwater landslide, we installed four numerical wave gauges located at $x\ =\ x_{\,0}\,$, $-50.0\,$, $0.0$ and $50.0\,$. The synthetic wave records are presented in Figure~\ref{fig:gauges}. One can see that the biggest quantity of primary interest is the wave run-up onto left and right beaches surrounding the fluid domain. This quantity is estimated numerically using the previously described algorithm. The shoreline motion is represented in Figure~\ref{fig:runup}. One can see that the landslide scenario under consideration produces much higher run-up values on the beach opposite to the slope where the sliding process takes place. Finally, in order to illustrate the energy transfer process from the landslide motion to the fluid layer, we show the evolution of both energies during the generation process in Figure~\ref{fig:energy}. We recall that the fluid potential, kinetic and total energies are defined correspondingly as
\begin{equation*}
  \Pi\,(t)\ \eqdef\ \frac12\;\int_\R g\,\eta^{\,2}\; \ud\,x\,, \quad
  \K\,(t)\ \eqdef\ \frac12\;\int_\R (d\ +\ \eta)\,u^{\,2}\; \ud\,x\,, \quad
  \E\,(t)\ \eqdef\ \Pi\,(t)\ +\ \K\,(t)\,.
\end{equation*}
The landslide kinetic energy is readily obtained from the differential Equation~\eqref{eq:ODE}:
\begin{equation*}
  \K_{\,\ell}\,(t)\ \eqdef\ \frac12\;(\gamma\ +\ c_{\,w})\,S\,\Bigl(\od{s}{t}\Bigr)^{\,2}\,.
\end{equation*}
Our computation shows that only about $10\%$ of the landslide energy is transmitted to the wave. This estimation is in complete accordance with values reported by \textsc{Harbitz} \etal \cite{Harbitz2006}.

\section{Discussion}\label{sec:concl}

Below we outline the main conclusions and perspectives of our study.


\subsection{Conclusions}

In the present study we revisited the celebrated \textsc{Peregrine} system for long waves propagation. Namely, our primary goal was to undertake a series of equivalent transformations which do not modify lower order dispersive terms $\O(\mu^{\,2})\,$, while extending the model stability and validity up to the shoreline. Moreover, the resulting governing equations possess an additional symmetry of the complete water wave problem which were broken as a result of the asymptotic expansion. Hence, our model remains invariant under the vertical translation (subgroup $G_{\,5}$ in Theorem~4.2, T.~\textsc{Benjamin} \& P.~\textsc{Olver} (1982) \cite{Benjamin1982}). The application of the invariantization process presented in this study can be extended to any other system of \textsc{Boussinesq} type. It can be viewed as a \emph{post treatment} procedure to be applied after the derivation of a particular model. The \textsc{Peregrine} system was chosen for illustrative purposes due to its importance and popularity in the water wave community. Of course, this system possesses also several nice properties which explain its wide usage in applications.

The developments made in this study are illustrated with several examples. First of all, we proposed an efficient numerical method to construct travelling wave solutions. Some comparisons with the classical Nonlinear Shallow Water equations (NSWE) were presented for the wave run-up problem onto a plane beach. The effect of dispersive terms is exemplified. In this study we also presented a model of a landslide motion over general curvilinear bottoms. This model takes into account the effects of bottom curvature, generally neglected in the literature \cite{Pelinovsky1996, Grilli1999, Watts2000, DiRisio2009}. Despite the inclusion of some new physical effects, the considered slide model is computationally inexpensive and can be potentially used in more operational context. We tested the m-\textsc{Peregrine} model on this more realistic case of the wave generation by an underwater landslide. The coupling with the m-\textsc{Peregrine} equations was done through the time-dependent bathymetry. Wave run-up records on non-flat beaches were computed. The proposed technique can be directly applied to perform a landslide hazard effects in real-world situations.


\subsection{Perspectives}

In the present manuscript we focused on the two-dimensional (2D) physical problem, which became a one-dimensional (1D) mathematical problem thanks to the elimination of explicit dependence on the vertical coordinate (1DH). In future works we are going to focus on the generalization of the m-\textsc{Peregrine} to the 2DH situation with two horizontal directions. There is another question which can be asked even in the 1D case --- the energy conservation issue. So far, a successful response to this question has been brought in the variational framework \cite{Clamond2015c}.


\subsection*{Acknowledgments}
\addcontentsline{toc}{subsection}{Acknowledgments}

D.~\textsc{Dutykh} \& A.~\textsc{Dur\'an} acknowledge the support from project MTM2014-54710-P entitled ``\textit{Numerical Analysis of Nonlinear Nonlocal Evolution Problems}'' (NANNEP). D.~\textsc{Mitsotakis} was supported by the \textsc{Marsden Fund} administered by the Royal Society of \textsc{New Zealand}.


\addcontentsline{toc}{section}{References}
\bibliographystyle{abbrv}
\bibliography{biblio}

\begin{thebibliography}{10}

\bibitem{Antuono2009}
M.~Antuono, V.~Y. Liapidevskii, and M.~Brocchini.
\newblock {Dispersive Nonlinear Shallow-Water Equations}.
\newblock {\em Studies in Applied Mathematics}, 122(1):1--28, 2009.

\bibitem{Audusse2004a}
E.~Audusse.
\newblock {\em {Mod{\'{e}}lisation hyperbolique et analyse num{\'{e}}rique pour
  les {\'{e}}coulements en eaux peu profondes}}.
\newblock PhD thesis, Universit{\'{e}} Paris {\{}VI{\}}, 2004.

\bibitem{Barth1994}
T.~J. Barth.
\newblock {Aspects of unstructured grids and finite-volume solvers for the
  Euler and Navier-Stokes equations}.
\newblock {\em Lecture series - van Karman Institute for Fluid Dynamics},
  5:1--140, 1994.

\bibitem{Barth2004}
T.~J. Barth and M.~Ohlberger.
\newblock {Finite Volume Methods: Foundation and Analysis}.
\newblock In E.~Stein, R.~de~Borst, and T.~J.~R. Hughes, editors, {\em
  Encyclopedia of Computational Mechanics}. John Wiley {\&} Sons, Ltd,
  Chichester, UK, nov 2004.

\bibitem{Batchelor2000}
G.~K. Batchelor.
\newblock {\em {An introduction to fluid dynamics}}, volume~61 of {\em
  Cambridge mathematical library}.
\newblock Cambridge University Press, 2000.

\bibitem{Beisel2012}
S.~A. Beisel, L.~B. Chubarov, D.~Dutykh, G.~S. Khakimzyanov, and N.~Y. Shokina.
\newblock {Simulation of surface waves generated by an underwater landslide in
  a bounded reservoir}.
\newblock {\em Russ. J. Numer. Anal. Math. Modelling}, 27(6):539--558, 2012.

\bibitem{Bellec2016}
S.~Bellec and M.~Colin.
\newblock {On the existence of solitary waves for Boussinesq type equations and
  Cauchy problem for a new conservative model}.
\newblock {\em Adv. Differential Equations}, 21(9/10):945--976, 2016.

\bibitem{Bellotti2001}
G.~Bellotti and M.~Brocchini.
\newblock {On the shoreline boundary conditions for Boussinesq-type models}.
\newblock {\em Int. J. Num. Meth. in Fluids}, 37(4):479--500, 2001.

\bibitem{Bellotti2002}
G.~Bellotti and M.~Brocchini.
\newblock {On using Boussinesq-type equations near the soreline: a note of
  caution}.
\newblock {\em Ocean Engineering}, 29:1569--1575, 2002.

\bibitem{Benjamin1982}
T.~B. Benjamin and P.~J. Olver.
\newblock {Hamiltonian structure, symmetries and conservation laws for water
  waves}.
\newblock {\em J. Fluid Mech}, 125:137--185, 1982.

\bibitem{Benkhaldoun2008}
F.~Benkhaldoun and M.~Sea{\"{i}}d.
\newblock {New finite-volume relaxation methods for the third-order
  differential equations}.
\newblock {\em Commun. Comput. Phys.}, 4:820--837, 2008.

\bibitem{Bernard2007}
E.~N. Bernard and V.~V. Titov.
\newblock {Improving tsunami forecast skill using deep ocean observations}.
\newblock {\em Mar. Technol. Soc. J.}, 40(4):23--26, 2007.

\bibitem{Bluman2010}
G.~W. Bluman, A.~F. Cheviakov, and S.~C. Anco.
\newblock {\em {Applications of Symmetry Methods to Partial Differential
  Equations}}.
\newblock Springer, New York, 2010.

\bibitem{Bogacki1989}
P.~Bogacki and L.~F. Shampine.
\newblock {A 3(2) pair of Runge-Kutta formulas}.
\newblock {\em Appl. Math. Lett.}, 2(4):321--325, 1989.

\bibitem{bouss}
J.~V. Boussinesq.
\newblock {Th{\'{e}}orie de l'intumescence liquide appel{\'{e}}e onde solitaire
  ou de translation se propageant dans un canal rectangulaire}.
\newblock {\em C.R. Acad. Sci. Paris S{\'{e}}r. A-B}, 72:755--759, 1871.

\bibitem{Boyd1986}
J.~P. Boyd.
\newblock {Solitons from sine waves: analytical and numerical methods for
  non-integrable solitary and cnoidal waves}.
\newblock {\em Physica D}, 21:227--246, 1986.

\bibitem{Boyd2000}
J.~P. Boyd.
\newblock {\em {Chebyshev and Fourier Spectral Methods}}.
\newblock New York, 2nd edition, 2000.

\bibitem{Boyd2002a}
J.~P. Boyd.
\newblock {A comparison of numerical algorithms for Fourier extension of the
  first, second and third kinds}.
\newblock {\em J. Comput. Phys.}, 178:118--160, 2002.

\bibitem{Boyd2002b}
J.~P. Boyd.
\newblock {Deleted residuals, the QR-factored Newton iteration, and other
  methods for formally overdetermined determinate discretizations of nonlinear
  eigenproblems for solitary, cnoidal, and shock waves}.
\newblock {\em J. Comput. Phys.}, 179:216--237, 2002.

\bibitem{Brocchini2013}
M.~Brocchini.
\newblock {A reasoned overview on Boussinesq-type models: the interplay between
  physics, mathematics and numerics}.
\newblock {\em Proc. R. Soc. A}, 469(2160):20130496, oct 2013.

\bibitem{Chambarel2009}
J.~Chambarel, C.~Kharif, and J.~Touboul.
\newblock {Head-on collision of two solitary waves and residual falling jet
  formation}.
\newblock {\em Nonlin. Processes Geophys.}, 16:111--122, 2009.

\bibitem{Cheviakov2010a}
A.~F. Cheviakov.
\newblock {Computation of fluxes of conservation laws}.
\newblock {\em J. Eng. Math.}, 66(1-3):153--173, mar 2010.

\bibitem{Christov2001}
C.~I. Christov.
\newblock {An energy-consistent dispersive shallow-water model}.
\newblock {\em Wave Motion}, 34:161--174, 2001.

\bibitem{Chubarov2011}
L.~B. Chubarov, G.~S. Khakimzyanov, and N.~Y. Shokina.
\newblock {Numerical modelling of surface water waves arising due to movement
  of underwater landslide on irregular bottom slope}.
\newblock In {\em Notes on Numerical Fluid Mechanics and Multidisciplinary
  Design: Computational Science and High Performance Computing IV}, pages
  75--91. Springer-Verlag, Berlin, Heidelberg, vol. 115 edition, 2011.

\bibitem{Clamond2009}
D.~Clamond and D.~Dutykh.
\newblock {Practical use of variational principles for modeling water waves}.
\newblock {\em Phys. D}, 241(1):25--36, 2012.

\bibitem{Clamond2015c}
D.~Clamond, D.~Dutykh, and D.~Mitsotakis.
\newblock {Conservative modified Serre--Green--Naghdi equations with improved
  dispersion characteristics}.
\newblock {\em Comm. Nonlin. Sci. Num. Sim.}, 45:245--257, 2017.

\bibitem{DeKaKa}
A.~I. Delis, M.~Kazolea, and N.~A. Kampanis.
\newblock {A robust high-resolution finite volume scheme for the simulation of
  long waves over complex domains}.
\newblock {\em Int. J. Numer. Meth. Fluids}, 56:419--452, 2008.

\bibitem{Demmel1997}
J.~W. Demmel.
\newblock {\em {Applied Numerical Linear Algebra}}.
\newblock SIAM, Philadelphia, 1997.

\bibitem{DiRisio2009}
M.~{Di Risio}, G.~Bellotti, A.~Panizzo, and P.~{De Girolamo}.
\newblock {Three-dimensional experiments on landslide generated waves at a
  sloping coast}.
\newblock {\em Coastal Engineering}, 56(5-6):659--671, 2009.

\bibitem{DDMM}
V.~A. Dougalis, A.~Dur{\'{a}}n, M.~A. Lopez-Marcos, and D.~E. Mitsotakis.
\newblock {A numerical study of the stability of solitary waves of Bona-Smith
  family of Boussinesq systems}.
\newblock {\em J. Nonlinear Sci.}, 17:595--607, 2007.

\bibitem{Duran2013}
A.~Duran, D.~Dutykh, and D.~Mitsotakis.
\newblock {On the Galilean Invariance of Some Nonlinear Dispersive Wave
  Equations}.
\newblock {\em Stud. Appl. Math.}, 131(4):359--388, nov 2013.

\bibitem{Dutykh2011a}
D.~Dutykh, D.~Clamond, P.~Milewski, and D.~Mitsotakis.
\newblock {Finite volume and pseudo-spectral schemes for the fully nonlinear 1D
  Serre equations}.
\newblock {\em Eur. J. Appl. Math.}, 24(05):761--787, 2013.

\bibitem{Dutykh2006}
D.~Dutykh and F.~Dias.
\newblock {Water waves generated by a moving bottom}.
\newblock In A.~Kundu, editor, {\em Tsunami and Nonlinear waves}, pages 65--96.
  Springer Verlag (Geo Sc.), 2007.

\bibitem{Dutykh2007b}
D.~Dutykh and F.~Dias.
\newblock {Tsunami generation by dynamic displacement of sea bed due to
  dip-slip faulting}.
\newblock {\em Mathematics and Computers in Simulation}, 80(4):837--848, 2009.

\bibitem{Dutykh2010}
D.~Dutykh and F.~Dias.
\newblock {Influence of sedimentary layering on tsunami generation}.
\newblock {\em Computer Methods in Applied Mechanics and Engineering},
  199(21-22):1268--1275, 2010.

\bibitem{Dutykh2011d}
D.~Dutykh and H.~Kalisch.
\newblock {Boussinesq modeling of surface waves due to underwater landslides}.
\newblock {\em Nonlin. Processes Geophys.}, 20(3):267--285, may 2013.

\bibitem{Dutykh2011e}
D.~Dutykh, T.~Katsaounis, and D.~Mitsotakis.
\newblock {Finite volume schemes for dispersive wave propagation and runup}.
\newblock {\em J. Comput. Phys.}, 230(8):3035--3061, apr 2011.

\bibitem{Dutykh2010c}
D.~Dutykh and D.~Mitsotakis.
\newblock {On the relevance of the dam break problem in the context of
  nonlinear shallow water equations}.
\newblock {\em Discrete and Continuous Dynamical Systems - Series B},
  13(4):799--818, 2010.

\bibitem{Dutykh2012}
D.~Dutykh, D.~Mitsotakis, S.~A. Beisel, and N.~Y. Shokina.
\newblock {Dispersive waves generated by an underwater landslide}.
\newblock In E.~Vazquez-Cendon, A.~Hidalgo, P.~Garcia-Navarro, and L.~Cea,
  editors, {\em Numerical Methods for Hyperbolic Equations: Theory and
  Applications}, pages 245--250. CRC Press, Boca Raton, London, New York,
  Leiden, 2013.

\bibitem{Dutykh2009a}
D.~Dutykh, R.~Poncet, and F.~Dias.
\newblock {The VOLNA code for the numerical modeling of tsunami waves:
  Generation, propagation and inundation}.
\newblock {\em Eur. J. Mech. B/Fluids}, 30(6):598--615, 2011.

\bibitem{EIK}
K.~S. Erduran, S.~Ilic, and V.~Kutija.
\newblock {Hybrid finite-volume finite-difference scheme for the solution of
  Boussinesq equations}.
\newblock {\em Int. J. Numer. Meth. Fluids}, 49:1213--1232, 2005.

\bibitem{Fenton1972}
J.~Fenton.
\newblock {A ninth-order solution for the solitary wave}.
\newblock {\em J. Fluid Mech}, 53(2):257--271, 1972.

\bibitem{Fernandez-Nieto2007}
E.~D. Fernandez-Nieto, F.~Bouchut, D.~Bresch, M.~J. Castro-Diaz, and
  A.~Mangeney.
\newblock {A new Savage-Hutter type models for submarine avalanches and
  generated tsunami}.
\newblock {\em J. Comput. Phys.}, 227(16):7720--7754, 2008.

\bibitem{Filippini2015}
A.~G. Filippini, S.~Bellec, M.~Colin, and M.~Ricchiuto.
\newblock {On the nonlinear behaviour of Boussinesq type models:
  Amplitude-velocity vs amplitude-flux forms}.
\newblock {\em Coastal Engineering}, 99:109--123, 2015.

\bibitem{Ghidaglia1996}
J.-M. Ghidaglia, A.~Kumbaro, and G.~{Le Coq}.
\newblock {Une m{\'{e}}thode volumes-finis {\`{a}} flux caract{\'{e}}ristiques
  pour la r{\'{e}}solution num{\'{e}}rique des syst{\`{e}}mes hyperboliques de
  lois de conservation}.
\newblock {\em C. R. Acad. Sci. I}, 322:981--988, 1996.

\bibitem{Golub1996}
G.~Golub and C.~{Van Loan}.
\newblock {\em {Matrix Computations}}.
\newblock J. Hopkins University Press, 3rd ed. edition, 1996.

\bibitem{Green1976}
A.~E. Green and P.~M. Naghdi.
\newblock {A derivation of equations for wave propagation in water of variable
  depth}.
\newblock {\em J. Fluid Mech.}, 78:237--246, 1976.

\bibitem{Grilli1999}
S.~T. Grilli and P.~Watts.
\newblock {Modeling of waves generated by a moving submerged body. Applications
  to underwater landslides}.
\newblock {\em Engineering Analysis with boundary elements}, 23:645--656, 1999.

\bibitem{Harbitz2006}
C.~B. Harbitz, F.~Lovholt, G.~Pedersen, S.~Glimsdal, and D.~G. Masson.
\newblock {Mechanisms of tsunami generation by submarine landslides - a short
  review}.
\newblock {\em Norwegian Journal of Geology}, 86(3):255--264, 2006.

\bibitem{Harten1989}
A.~Harten.
\newblock {ENO schemes with subcell resolution}.
\newblock {\em J. Comput. Phys}, 83:148--184, 1989.

\bibitem{HaOs}
A.~Harten and S.~Osher.
\newblock {Uniformly high-order accurate nonscillatory schemes. I}.
\newblock {\em SIAM J. Numer. Anal.}, 24:279--309, 1987.

\bibitem{Higham2002}
N.~J. Higham.
\newblock {\em {Accuracy and Stability of Numerical Algorithms}}.
\newblock SIAM Philadelphia, 2nd ed. edition, 2002.

\bibitem{Kalisch2004}
H.~Kalisch.
\newblock {Stability of solitary waves for a nonlinearly dispersive equation}.
\newblock {\em Discrete and Continuous Dynamical Systems}, 10:709--717, 2004.

\bibitem{Khakimzyanov2016a}
G.~S. Khakimzyanov, D.~Dutykh, and Z.~I. Fedotova.
\newblock {Dispersive shallow water wave modelling. Part III: Model derivation
  on a globally spherical geometry}.
\newblock {\em Commun. Comput. Phys.}, 23(2):315--360, 2018.

\bibitem{Khakimzyanov2016c}
G.~S. Khakimzyanov, D.~Dutykh, Z.~I. Fedotova, and D.~E. Mitsotakis.
\newblock {Dispersive shallow water wave modelling. Part I: Model derivation on
  a globally flat space}.
\newblock {\em Commun. Comput. Phys.}, 23(1):1--29, 2018.

\bibitem{Kolgan1975}
N.~E. Kolgan.
\newblock {Finite-difference schemes for computation of three dimensional
  solutions of gas dynamics and calculation of a flow over a body under an
  angle of attack}.
\newblock {\em Uchenye Zapiski TsaGI [Sci. Notes Central Inst. Aerodyn]},
  6(2):1--6, 1975.

\bibitem{MR2011541}
P.~L.-F. Liu, P.~Lynett, and C.~E. Synolakis.
\newblock {Analytical solutions for forced long waves on a sloping beach}.
\newblock {\em J. Fluid Mech.}, 478:101--109, 2003.

\bibitem{Longuet-Higgins1974}
M.~S. Longuet-Higgins and J.~Fenton.
\newblock {On the Mass, Momentum, Energy and Circulation of a Solitary Wave.
  II}.
\newblock {\em Proc. R. Soc. A}, 340(1623):471--493, 1974.

\bibitem{LordRayleigh1876}
J.~W.~S. {Lord Rayleigh}.
\newblock {On Waves}.
\newblock {\em Phil. Mag.}, 1:257--279, 1876.

\bibitem{Madsen03}
P.~A. Madsen, H.~B. Bingham, and H.~A. Schaffer.
\newblock {Boussinesq-type formulations for fully nonlinear and extremely
  dispersive water waves: derivation and analysis}.
\newblock {\em Proc. R. Soc. Lond. A}, 459:1075--1104, 2003.

\bibitem{Madsen1999}
P.~A. Madsen and H.~A. Schaffer.
\newblock {A review of Boussinesq-type equations for surface gravity waves}.
\newblock {\em Adv. Coastal Ocean Engng}, 5:1--94, 1999.

\bibitem{Madsen1997}
P.~A. Madsen, H.~A. Sorensen, and H.~A. Schaffer.
\newblock {Surf zone dynamics simulated by a Boussinesq-type model. Part I.
  Model description and cross-shore motion of regular waves}.
\newblock {\em Coastal Engineering}, 32:255--287, 1997.

\bibitem{Nwogu1993}
O.~Nwogu.
\newblock {Alternative form of Boussinesq equations for nearshore wave
  propagation}.
\newblock {\em J. Waterway, Port, Coastal and Ocean Engineering}, 119:618--638,
  1993.

\bibitem{Okal1988}
E.~A. Okal.
\newblock {Seismic Parameters Controlling Far-field Tsunami Amplitudes: A
  Review}.
\newblock {\em Natural Hazards}, 1:67--96, 1988.

\bibitem{Okal2003}
E.~A. Okal and C.~E. Synolakis.
\newblock {A theoretical comparison of tsunamis from dislocations and
  landslides}.
\newblock {\em Pure and Applied Geophysics}, 160:2177--2188, 2003.

\bibitem{Okal2004}
E.~A. Okal and C.~E. Synolakis.
\newblock {Source discriminants for near-field tsunamis}.
\newblock {\em Geophys. J. Int.}, 158:899--912, 2004.

\bibitem{Olver1993}
P.~J. Olver.
\newblock {\em {Applications of Lie groups to differential equations}}, volume
  107 (2nd e of {\em Graduate Texts in Mathematics}.
\newblock Springer-Verlag, 1993.

\bibitem{Pascal2002}
F.~Pascal.
\newblock {\em {Sur des m{\'{e}}thodes d'approximation effectives et d'analyse
  num{\'{e}}rique pour les {\'{e}}quations de la m{\'{e}}canique de fluides}}.
\newblock Habilitation {\`{a}} diriger des recherches, Universit{\'{e}} de
  Paris-Sud, 2002.

\bibitem{Pelinovsky1996}
E.~Pelinovsky and A.~Poplavsky.
\newblock {Simplified model of tsunami generation by submarine landslides}.
\newblock {\em Physics and Chemistry of the Earth}, 21(12):13--17, 1996.

\bibitem{Peregrine1967}
D.~H. Peregrine.
\newblock {Long waves on a beach}.
\newblock {\em J. Fluid Mech.}, 27:815--827, 1967.

\bibitem{Sandee1991}
J.~Sandee and K.~Hutter.
\newblock {On the development of the theory of the solitary wave. A historical
  essay}.
\newblock {\em Acta Mechanica}, 86:111--152, 1991.

\bibitem{Serre1953}
F.~Serre.
\newblock {Contribution {\`{a}} l'{\'{e}}tude des {\'{e}}coulements permanents
  et variables dans les canaux}.
\newblock {\em La Houille blanche}, 8:374--388, 1953.

\bibitem{Shampine1997}
L.~F. Shampine and M.~W. Reichelt.
\newblock {The MATLAB ODE Suite}.
\newblock {\em SIAM J. Sci. Comput.}, 18:1--22, 1997.

\bibitem{Soderlind2003}
G.~S{\"{o}}derlind.
\newblock {Digital filters in adaptive time-stepping}.
\newblock {\em ACM Trans. Math. Software}, 29:1--26, 2003.

\bibitem{Soderlind2006}
G.~S{\"{o}}derlind and L.~Wang.
\newblock {Adaptive time-stepping and computational stability}.
\newblock {\em J. Comp. Appl. Math.}, 185(2):225--243, 2006.

\bibitem{Synolakis1987}
C.~E. Synolakis.
\newblock {The runup of solitary waves}.
\newblock {\em J. Fluid Mech.}, 185:523--545, 1987.

\bibitem{Syno2006}
C.~E. Synolakis and E.~N. Bernard.
\newblock {Tsunami science before and beyond Boxing Day 2004}.
\newblock {\em Phil. Trans. R. Soc. A}, 364:2231--2265, 2006.

\bibitem{Tinti2001}
S.~Tinti, E.~Bortolucci, and C.~Chiavettieri.
\newblock {Tsunami Excitation by Submarine Slides in Shallow-water
  Approximation}.
\newblock {\em Pure appl. geophys.}, 158:759--797, 2001.

\bibitem{Titov2005}
V.~V. Titov, F.~I. Gonzalez, E.~N. Bernard, M.~C. Eble, H.~O. Mofjeld, J.~C.
  Newman, and A.~J. Venturato.
\newblock {Real-Time Tsunami Forecasting: Challenges and Solutions}.
\newblock {\em Natural Hazards}, 35:41--58, 2005.

\bibitem{todo2}
M.~I. Todorovska, A.~Hayir, and M.~D. Trifunac.
\newblock {A note on tsunami amplitudes above submarine slides and slumps}.
\newblock {\em Soil Dynamics and Earthquake Engineering}, 22:129--141, 2002.

\bibitem{Toro2009}
E.~F. Toro.
\newblock {\em {Riemann Solvers and Numerical Methods for Fluid Dynamics}}.
\newblock Springer, Berlin, Heidelberg, 2009.

\bibitem{Ursell1953}
F.~Ursell.
\newblock {The long-wave paradox in the theory of gravity waves}.
\newblock {\em Proc. Camb. Phil. Soc.}, 49:685--694, 1953.

\bibitem{Leer1979}
B.~van Leer.
\newblock {Towards the ultimate conservative difference scheme V: a second
  order sequel to Godunov' method}.
\newblock {\em J. Comput. Phys.}, 32:101--136, 1979.

\bibitem{Leer2006}
B.~van Leer.
\newblock {Upwind and High-Resolution Methods for Compressible Flow: From Donor
  Cell to Residual-Distribution Schemes}.
\newblock {\em Commun. Comput. Phys.}, 1:192--206, 2006.

\bibitem{Watts2000}
P.~Watts, F.~Imamura, and S.~T. Grilli.
\newblock {Comparing model simulations of three benchmark tsunami generation
  cases}.
\newblock {\em Science of Tsunami Hazards}, 18(2):107--123, 2000.

\bibitem{Wu1981}
T.~Y. Wu.
\newblock {Long Waves in Ocean and Coastal Waters}.
\newblock {\em Journal of Engineering Mechanics}, 107:501--522, 1981.

\bibitem{Wu1987}
T.~Y.~T. Wu.
\newblock {Generation of upstream advancing solitons by moving disturbances}.
\newblock {\em J. Fluid Mech.}, 184:75--99, 1987.

\bibitem{Xing2005}
Y.~Xing and C.-W. Shu.
\newblock {High order finite difference WENO schemes with the exact
  conservation property for the shallow water equations}.
\newblock {\em J. Comput. Phys.}, 208:206--227, 2005.

\bibitem{Yang2010}
J.~Yang.
\newblock {\em {Nonlinear Waves in Integrable and Nonintegrable Systems}}.
\newblock Society for Industrial and Applied Mathematics, Philadelphia, jan
  2010.

\bibitem{Zelt1991}
J.~A. Zelt.
\newblock {The run-up of nonbreaking and breaking solitary waves}.
\newblock {\em Coastal Engineering}, 15:205--246, 1991.

\bibitem{Zhou2002}
J.~G. Zhou, D.~M. Causon, D.~M. Ingram, and C.~G. Mingham.
\newblock {Numerical solutions of the shallow water equations with
  discontinuous bed topography}.
\newblock {\em Int. J. Numer. Meth. Fluids}, 38:769--788, 2002.

\end{thebibliography}
\bigskip

\end{document}